\documentclass{article}
\usepackage{graphicx,amssymb}
\usepackage[rightcaption]{sidecap}
\usepackage{natbib}
\usepackage[margin=0.5in]{geometry}
\usepackage{color}
\usepackage{enumitem}
\usepackage{authblk}
\usepackage{hyperref}

\title{The JWST Galactic Center Survey \\
A White Paper}

\author[1]{Rainer Sch{\"o}del}
\author[2,3]{Steve Longmore}
\author[2]{Jonny Henshaw}
\author[4]{Adam Ginsburg}
\author[5]{John Bally}
\author[6]{Anja Feldmeier}
\author[7]{Matt Hosek}
\author[8]{Francisco Nogueras Lara}
\author[7]{Anna Ciurlo}
\author[9,3]{M\'elanie Chevance}
% ...
\affil[1]{Instituto de Astrof\'isica de Andaluc\'ia (CSIC), Granada, Spain}
\affil[2]{Liverpool John Moores University, Liverpool, UK}
\affil[3]{ Cosmic Origins Of Life (COOL) Research DAO, \url{coolresearch.io}}
\affil[4]{University of Florida, Gainesville, FL, USA}
\affil[5]{University of Colorado Boulder, Boulder, CO, USA}
\affil[6]{Max-Planck-Institute for Astronomy, Heidelberg, Germany}
\affil[7]{University of California Los Angeles, Los Angeles,  CA, USA}
\affil[8]{European Southern Observatory, Garching, Germany}
\affil[9]{University of Heidelberg, Heidelberg, Germany}
% ...

\date{\today}

\newcommand{\msol}{M$_{\odot}$}
\newcommand{\Msol}{M$_{\odot}$}

\newcommand{\kms}{km~s$\rm ^{-1}$}
\newcommand{\Htwo}{H$\rm {_2}$}
\newcommand{\SFR}{$\rm M_{\odot} yr^{-1}$}
\newcommand{\sgra}{Sgr\,A*}

\newcommand*\arcsec{\ensuremath{^{\prime\prime}}}

\begin{document}

\maketitle

\newpage
\section{Abstract}

The inner hundred parsecs of the Milky Way hosts the nearest supermassive black hole, largest reservoir of dense gas, greatest stellar density, hundreds of massive main and post main sequence stars, and the highest volume density of supernovae in the Galaxy. As the nearest environment in which it is possible to simultaneously observe many of the extreme processes shaping the Universe, it is one of the most well-studied regions in astrophysics. Due to its proximity, we can study the center of our Galaxy on scales down to a few hundred AU, a hundred times better than in similar Local Group galaxies and thousands of times better than in the nearest active galaxies. The Galactic Center is therefore of outstanding astrophysical interest. However, in spite of intense observational work over the past decades, there are still fundamental things unknown about the Galactic Center, because it is an extremely challenging region to observe. JWST has the unique capability to provide us with the necessary, game-changing data. In this White Paper, we advocate for a JWST NIRCam survey that aims at solving central questions. As a community, we have identified the key unknowns that are limiting the potential of the Galactic Center as a laboratory for extreme astrophysics and understanding how galactic nuclei shape the galaxy population: i) the 3D structure and kinematics of gas and stars; ii)  ancient star formation  and its relation with the overall  history of the Milky Way, as well as recent star formation and its implications for the overall energetics of our galaxy's nucleus; and iii) the (non-)universality of star formation and the stellar initial mass function. We advocate for a large-area, multi-epoch, multi-wavelength NIRCam survey of the inner 100\,pc of the Galaxy in the form of a Treasury GO JWST Large Program that is open to the community. We describe how this survey will derive the physical and kinematic properties of $\sim$10,000,000 stars, how this will solve the key unknowns and provide a valuable resource for the community with long-lasting legacy value.

\section{Community Science Goals and Recommendations}

Given the breadth of science topics covered by Galactic Center research, we have undertaken an open community consultation process in order to identify the highest priority science questions and determine how these can be addressed with JWST. We begin with a concise summary of the major open questions to be addressed and the unique capabilities of JWST which make it revolutionary for Galactic Center research. We then outline the survey requirements needed to directly address these key open questions and the synergy of these observations with current and future facilities.

\subsection{Key open questions in Galactic Center research}\label{questions}

The JWST Galactic Center (GC) Survey will tackle major open questions in the field: 

\begin{enumerate}[leftmargin=*]
        \item What is the formation history of the Galactic Center and its relation to the overall formation history of the Milky Way?
        \item How much stellar mass formed in the past $\sim$30\,Myr and what does this imply for the overall energetics of the GC?
        \item What is the origin of, and environmental variation in, the stellar initial mass function?
        \item Why is the star formation rate one to two orders of magnitude lower than predicted by standard star-formation-dense-gas relations?
        \item What is the 3D structure of the interstellar medium (ISM) orbiting and fueling accretion and star formation at the Galactic Center?
    \end{enumerate}

\noindent We expand on each of these questions in \S\ref{science}.\\
 
 \noindent By being able to resolve physical processes down to size scales separating individual stars, the survey will provide a foundation for addressing key open questions in other fields: What drives the mass flows and energy cycles in extragalactic nuclei and high-z environments? What shapes star formation and the evolution of nuclear star clusters, nuclear stellar discs and their interaction with central black holes? In what way are astrophysical processes different in extreme environments?

\begin{figure}[!b]
\includegraphics[width=\textwidth]{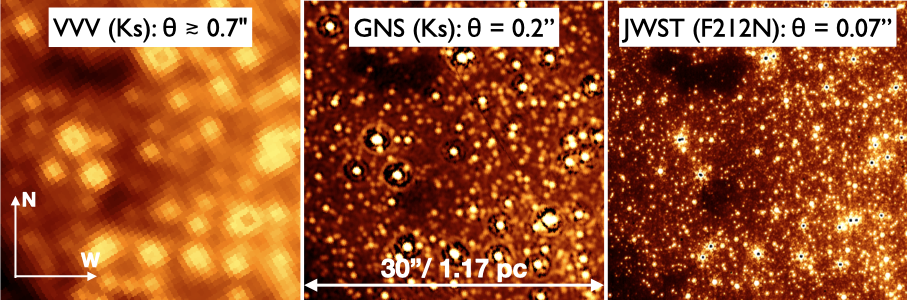}
\caption{Comparison of the same field in the GC, about $25"$ northeast of \sgra, imaged by
  VVV \citep{Minniti:2010fk}, GALACTICNUCLEUS \citep{Nogueras-Lara:2019yj} and JWST (JWS Proposal 1939, PI J. Lu). These images illustrate why angular resolution is of key importance for studying the Galactic Center, where crowding is a serious limitation. The angular resolution increases roughly by a factor of three between the images, meaning that source confusion is reduced by a factor of almost ten between the images obtained by the different instruments. \label{fig:comparison}}
\end{figure}

\subsection{The unique observational capabilities of JWST in Galactic Center context}\label{capabilities}

JWST has unique capabilities that make it possible to solve the open questions in Galactic Center (GC) science listed in $\S$\ref{questions}:

\begin{itemize}[leftmargin=*]
        \item {\bf Wavelength coverage}: Interstellar extinction towards the GC ranges from $\gtrsim30$\,mag to $>100$\,mag in the visual regime. It decreases steeply in the near-infrared and reaches a minimum of $\sim$1\,mag at $\lambda\approx5\,\mu$m. NIRCam 
        is on the order of $10^{4}$ times more sensitive than any ground-based instrument at this wavelength. JWST/NIRCam  is therefore unique and essential for studying the stellar structure, stellar population, and young stellar objects at the GC.
        \item {\bf Angular resolution}: Crowding is the main factor that limits the detection of faint stars at the GC, such as old stars near the main sequence turn-off and low-mass YSOs (Fig.\,\ref{fig:comparison}). JWST provides a ten times higher angular resolution than Spitzer and can deliver almost the resolution of adaptive optics-assisted cameras at ground-based 8~m-class telescopes. However, since precision measurements with the latter are strongly limited by anisoplanatic effects, their useful field-of-view is only approximately 0.25\,arcmin$^{2}$, while the effective field-of-view of NIRCam is roughly 40 times larger in each channel. 
        \item {\bf Sensitivity}: The factor $10^{4}$  improvement in sensitivity with NIRCam/JWST will allow us to observe faint, deeply embedded stars and young stellar objects (YSOs) in the GC that cannot be detected by any other space or ground-based telescope. JWST's longer wavelength coverage pierces further into the dusty regions and enables dereddening.
        \item {\bf The extremely stable PSF and optical system} of NIRCam/JWST delivers high-precision photometry and astrometry, enabling a wide range of multi-epoch science. This stability over a very large field-of-view is a major advantage over ground-based instruments. 
    \end{itemize}

Ground-based, adaptive optics-assisted instruments on 10m-class telescopes are no alternative to JWST, because (1) they can cover at most arcminute-wide fields with a systematically changing PSF due to anisoplanatic effects, (2) zero point variations, mostly due to PSF instability, are a problem for multi-epoch observations, and (3) they are about $10^4$ times less sensitive at $\lambda=5\,\mu$m than JWST. (1) implies that  accurate photometry and astrometry can only be performed over very small fields ($\sim$0.25\,arcmin$^{2}$), thus making any observations of large fields highly time-consuming. (3) means that less than 10\% of the stars detected at $\lambda=2.2\,\mu$m can be detected at $\lambda=4-5\,\mu$m from the ground. 

Even with NIRCam/JWST our main limitation in GC observations will be crowding, but the high angular resolution and sensitivity of NIRCam cannot be matched by any other instrument. A survey of the GC with NIRCam/JWST would be complete down to magnitudes of 20-21 in the near- and 18-19 in the mid-infrared, that is two ($\lambda=2\,\mu$m) to five ($\lambda=5\,\mu$m) magnitudes more sensitive than from the ground. Assuming a plausible luminosity function (see Fig.\,\ref{fig:obs}), this implies six (at $\lambda=2\,\mu$m) times more stars per area than in GALACTICNUCLEUS, the so far most complete GC survey \citep{Nogueras-Lara:2019yj},  to 40 times the number of stars that can be observed from the ground at $\lambda=4-5\,\mu$m (assuming adaptive optics observations at the VLT). {\bf A survey of the Galactic Center at $\lambda\approx5\,\mu$m  with sufficient sensitivity to achieve the key science goals is therefore all but impossible from the ground.}

\subsection{JWST will produce transformational measurements}\label{measurements}

The unique observational capabilities of JWST outlined in \S\ref{capabilities} will enable a number of new measurements that will provide transformational data for addressing each of the key science questions: 


\begin{itemize}[leftmargin=*]
        \item {\it The superb angular resolution of JWST's NIRCam is key to understanding the stellar population at the GC and its formation history (Fig.\,\ref{fig:obs}).} The detection of stars at the GC is limited primarily by source confusion.  Due to its angular resolution, NIRCam imaging will be about ten times less confused than the best existing Galactic Center survey (GALACTICNUCLEUS) and therefore reach about 2 magnitudes deeper. Assuming a mean extinction of 2\,mag at $\lambda=2\,\mu$m NIRCam will therefore detect main-sequence stars down to solar masses. This will allow us to detect the main sequence turnoff of populations as old as 10\,Gyr. {\it The de-reddened luminosity functions of NIRCam will therefore trace different star formation events with unprecedented clarity and will constrain the star formation history and its spatial change across the Galactic Center}.
        \item {\it Combined near- and mid-infrared observations will enable dereddening of individual stars}. The detectable stars have negligible intrinsic colors with the appropriate combination of near infrared (NIR) and mid-infrared (MIR) filters, so NIRCam has the unique capability to directly measure extinction toward each star. Currently, we rely on extinction maps with a spatial resolution of several arcseconds. Additionally, these maps are derived from averaging over many stars in a region. Precise dereddening is key to obtain reliable color-magnitude diagrams and luminosity functions that will serve to constrain the star formation history and to search for young, massive stars. Reddening measurements of individual stars will also be used to measure the column density  of dust along the line of sight.
        \item {\it With its high sensitivity JWST can potentially detect YSOs down to  L$_{bol}$ $\ge$0.1\,$L_{\odot}$} in fields where detection completeness is not limited by crowding, such as star-forming regions associated with dark clouds. Current data, mainly from the Spitzer Space Telescope, limit detections to massive YSOs, which are rare and, in the Spitzer data, confused with other sources.
        \item {\it NIRCam imaging is key to  distinguishing young, hot main sequence stars from red giant stars and to identifying candidate YSOs in color-magnitude diagrams.} This is currently not possible due to the uncertain  extinction correction (see above) or would require expensive spectroscopic observations. YSOs can be identified via their mid-infrared excess. Distinguishing hot main sequence stars from red giants will require accurate reddening measurements combined with short-wavelength imaging. The former requires JWST and the latter can be done with HST or from the ground. 
        \item {\it With a minimum of two NIRCam epochs, we can measure the proper motions of an estimated 10 million stars at the GC.}  No existing or planned mission (including Gaia NIR) is or will be able to achieve this.  A five year time base will allow proper motion measurements with an accuracy of at least $0.3$~\,mas\,yr$^{-1}$ (10\,km\,s$^{-1}$). Here we have considered $F212N=18$\,mag stars and have assumed that alignment between two observing epochs can be done with an accuracy of 1\,mas.  The proper motions will link the Galactic Center to the rest of the Milky Way where proper motions are measured with Gaia and the Vera Rubin observatory. Proper motions are key to disentangling different stellar structures (bar, nuclear stellar disk, nuclear star cluster), identifying accretion events onto the nuclear star cluster and understanding the structure of the nuclear stellar disk (is it an inner bar?).  Even a single epoch JWST survey will already yield about one million proper motions by combination with archival HST \citep[e.g.][]{Dong:2011ff,Libralato:2021qa} and ground-based images (from the GALACTICNUCLEUS survey). 
        
    \end{itemize}

\subsection{Survey requirements to address the key open questions}\label{requirements}

\begin{figure}[!b]
\includegraphics[width=\textwidth]{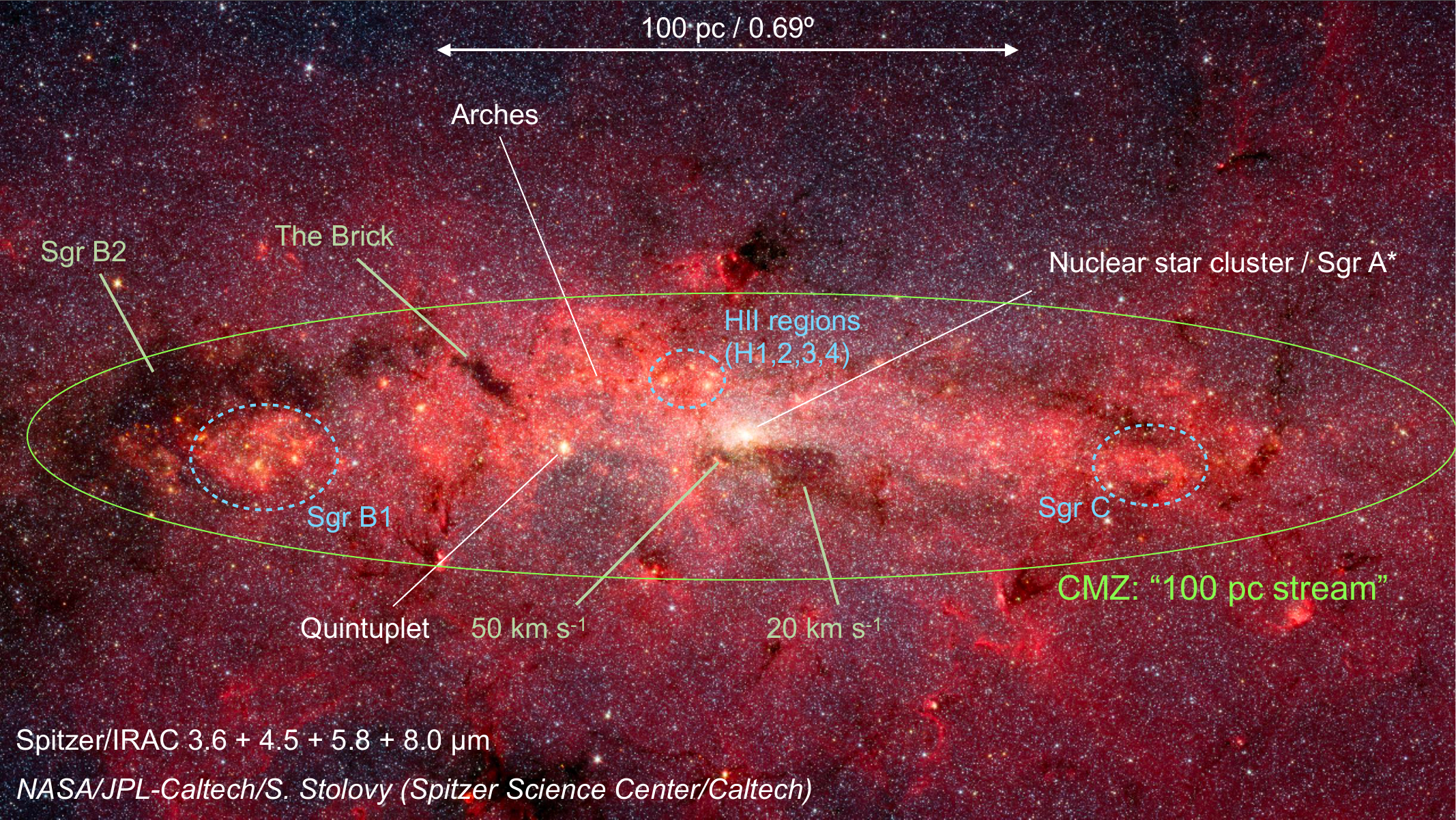}
\caption{Overview of the Galactic Center. Galactic north is up and
  east to the left. Prominent H\,II regions are labelled in blue, star
  clusters in white, and major molecular clouds in green. The large ellipse
  outlines the region containing most of the stellar mass of the NSD, most of the molecular gas in the CMZ, and the region of active star formation. \label{fig:overview}}
\end{figure}

Realising the science objectives in $\S$\ref{questions} requires a survey with the following properties:

    \begin{itemize}[leftmargin=*]
        \item Areal coverage: nuclear stellar disk and associated giant molecular clouds in the central molecular zone (about $1.25^{\circ}\times0.25^{\circ}$ or $180\,\mathrm{pc}\times36\,\mathrm{pc}$, Fig.\,\ref{fig:overview}).
        \item Filters: F140M, F187N, F212N, F405N and F480M to enable accurate dereddening, minimize saturation, trace ISM features, and obtain the best possible color-magnitude diagrams (CMDs). Since confusion is less of an issue at the shortest wavelengths, where extinction is significantly higher, an alternative to F140M imaging with NIRCam may be F127M imaging with WFC3/HST.
        \item Observing cadence: for accurate proper motion measurements we require observations at three epochs separated by 1, 5, and 10 years.
    \end{itemize}

\noindent These observations will provide the transformational measurements outlined in $\S$\ref{measurements} across the inner 100\, pc of the Galaxy. 
Each epoch of imaging will require about 137.5\,h of charged time (see \S\,\ref{sec:strategy}).

\subsection{Synergy with other facilities} 

A large-area, multi-epoch, multi-wavelength survey of the inner 100\,pc of the Galaxy in the form of a Treasury GO JWST Large Program is naturally synergistic with observations from other major facilities:

    \begin{itemize}[leftmargin=*]
        \item ALMA has recently completed the ALMA CMZ Exploration Survey (ACES) in the 3.2 mm continuum and over a dozen spectral lines with $\sim$2\arcsec\ angular resolution. 
        \item The HST has observed the nuclear star cluster and the Arches and Quintuplet clusters  during several epochs \citep[HST Proposal ID 12663, PI T. Do, ][]{Hosek:2022om}. The epoch 2009 HST Paschen$-\alpha$ survey and the proper motion work by \citet{Libralato:2021qa} cover  significant parts of the GC \citep{Dong:2011ff}. These data can be combined with the proposed JWST observations to obtain precision photometry and astrometry (proper motions) for  a few $10^5$ stars.
        \item SKA and ngVLA centimeter-wavelength observations of the GC will provide radio hydrogen recombination lines. Combined with infrared recombination lines from the JWST, this will enable measurement of the absolute extinction towards all ISM features. The high angular resolution and sensitivity of the JWST will also allow the identification of stellar counterparts to sources dominated by non-thermal emission such as pulsars and X-ray binaries. 
        \item The Roman Space Telescope (launch 2027) offers a compelling opportunity to perform a high-cadence survey of the GC region at HST-like resolution \citep[e.g.][]{Terry:2023lg} for time-domain science and the planne  \href{https://roman.gsfc.nasa.gov/science/galactic_plane_survey_definition.html}{Galactic Plane Survey} will image the GC with multiple filters. Because of its worse angular resolution Roman WFI images will hit the crowding limit at brighter magnitudes than JWST NIRCam.  The proposed JWST observations will greatly enhance the Roman data by providing star-by-star extinction measurements, much-improved depth in highly crowded and/or extinguished regions, identification of close/blended stars, whose photometry/astrometry will be significantly biased, and an epoch of high-precision astrometry for proper motion measurements.
        \item The ESO ELT (start of operations in 2028) will need JWST as a path finder to select fields for follow-up.
        \item Vera C. Rubin Observatory/LSST (start of operations in 2024): 
         Rubin will measure the proper motions of an estimated 200 million stars in the Milky Way. However, the high extinction means the optical telescope is effectively blind at the Galactic Center. JWST's sensitivity and wavelength coverage will enable accurate proper motions of 10 million stars towards the Galactic Center. 
         \item X-ray observatories have revealed the presence of an outflow from the Galactic Center \citep[e.g.][]{Ponti:2019ul,Predehl:2020py}. The proposed observations may allow us to connect star formation regions to outflows and thus verify the interpretation of X-ray observations.
         \item The future JASMINE infrared astrometry mission is focused on the Galactic Center and will provide us with ultra-high precision absolute proper motions for stars brighter than $K<14$\,mag in the GC \citep{Gouda:2018qe}.  The proposed JWST survey will allow us to quantitatively estimate astrometric biases of bright stars that are confused  with fainter ones. This is an important aspect for JASMINE and also the future Gaia NIR mission (see https://www.astro.lu.se/GaiaNIR).
    \end{itemize}

\noindent Together these surveys 
herald a revolution in the interpretation of current/future data, bring together research in different sub-fields, and answer key open science questions with enormous legacy potential.

\section{Solving multiple key science problems with a single data set}\label{science}

Figure\,\ref{fig:overview} provides an overview of the main structures in the Galactic Center (GC): the central molecular zone (CMZ), the nuclear stellar disk (NSD), and the nuclear star cluster (NSC). 

The $\sim$500$\times$50~pc CMZ contains 2 - 7$\times 10^7$ \Msol\ of dense molecular gas, corresponding to  $3-10$\%  of our Galaxy’s molecular gas. Most of the mass is
distributed in a roughly ring-like/elliptical shape of $\sim$100\,pc
($0.7^{\circ}$) radius around \sgra, the so-called 100\,pc-stream
\citep[Fig.\,\ref{fig:overview}, e.g.][]{Henshaw:2022nm}. The CMZ includes exceptional ISM features such as the $10^5$~\Msol\ ``Brick" \citep{Longmore:2012uq} and the $\sim 10^6$~\Msol\ Sgr\,B2 cloud \citep[][and references therein]{de-Pree:1995kl,Schmiedeke:2016ph,Ginsburg:2018oz}.  The CMZ occupies
$<0.005\%$ of the volume of the Milky Way's disk and reaches a mean molecular gas
density $>100$ times higher than in the Galactic disk. 

The NSD  consists of $\sim 10^9$ \Msol\ of stars in a flat, rotating, disk-like structure that has an effective radius of  $\sim$100\,pc and a scale-height of $\sim$40\,pc \citep{Launhardt:2002nx,Schonrich:2015uq,Henshaw:2022nm,Sormani:2022dq}.  
With a mass of $\sim2.5\times10^{7}$\,\msol\ and an effective radius of about 4\,pc, the NSC is the densest concentration of stars in the Galaxy \citep{Schodel:2014bn,Feldmeier-Krause:2017rt,Neumayer:2020jw}. The majority of the stars in the NSD and NSC, $\gtrsim80\%$, appear to have formed more than 8-10\,Gyr ago \citep{Nogueras-Lara:2020pp,Schodel:2020qc} and are the most metal-rich population of old stars in the Galaxy \citep{Schultheis:2021du}. At the kinematic and density center of the NSC lies the $4\times10^{6}$\,\msol\ super-massive black hole Sagittarius\,A* (\sgra)  \citep{Ghez:2008fk,Gravity-Collaboration:2020oy}.

 Averaged by volume, the mean star formation rate at the GC is one to two orders of magnitude higher than in the Galactic disk, with extreme conditions that resemble those in high redshift star-forming galaxies \citep{Kruijssen:2013tq}. Ongoing and intense past star formation activity is witnessed by  a broad range of observational evidence: Studies of classical Cepheids and of
the near-infrared luminosity function indicate that about
$1\times10^{6}$ \msol\ of stars formed in the GC in the past few tens
of Myr \citep{Matsunaga:2011uq,Nogueras-Lara:2020pp}. The GC contains three massive
($\sim$$10^{4}$\,\msol) young (2.5-6\,Myr) clusters: the Arches and
Quintuplet and the central parsec cluster
\citep[e.g.][]{Figer:1999fk,Bartko:2009fq,Lu:2013fk,Hosek:2022om}.  There are H\,II
regions and massive YSOs distributed throughout the nuclear disk
\citep[e.g.][]{Nandakumar:2018rw,Hankins:2019sw}. In addition, on the order of 100 apparently isolated massive young stars have been discovered distributed throughout the GC \citep{Dong:2011ff,Clark:2021fj}. 
There are about $10^{5}$\,\msol\ of young stars in the Sgr\,B1 H\,II region \citep{Nogueras-Lara:2022jz}.
Sgr B2 is currently the most active site of star formation in the Galaxy. It contains several hundred, highly embedded massive OB stars \citep[e.g.,][]{Schmiedeke:2016ph,Ginsburg:2018oz}.


\subsection{What is the formation history of the Galactic Center and its relation to the overall formation history of the Milky Way?}

The star formation history of the GC is essential for understanding two aspects of the evolution of the Milky Way. {\bf (1) The oldest stars in the nuclear stellar disk tell us the age of our Galaxy's bar, which funnels significant amounts of gas towards the center} \citep[e.g.][]{Baba:2019wg}. {\bf (2) Since black hole growth requires substantial gas infall, which drives star formation, the age structure of the stellar population will be correlated with the growth history of the central supermassive black hole, Sagittarius\,A*.} 

Interstellar extinction towards the GC is extremely high and rises steeply toward the optical. Therefore sensitive observations of its stellar population are limited to wavelengths $2-5\,\mu$m. Since the spread of intrinsic stellar colors of $>95\%$ of the detectable stars is $\lesssim 0.1$\,mag at these wavelengths, color-magnitude diagrams provide only very limited information. The best possible scenario is using short NIR observations, e.g.\ with the F140M filter, where the spread of intrinsic colors is the largest. 

\begin{SCfigure}[1.0][h]
\includegraphics[width=.4\textwidth]{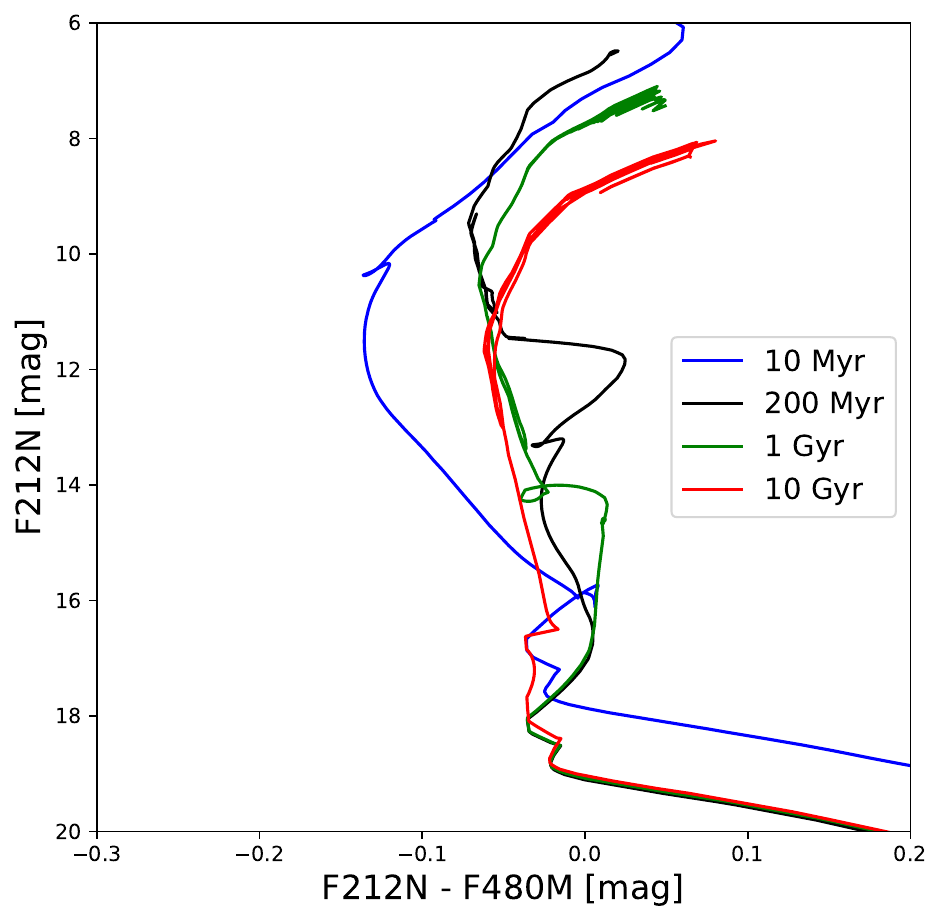}
\caption{Left: The intrinsic stellar colors at $F212N-F480M$  have a very small, roughly constant value for almost all stellar ages and a wide range of magnitudes. These NIRCam filters can therefore provide for an accurate measurement of extinction.  \label{fig:stellar_colors}}
\end{SCfigure}

\begin{figure}[p]
\centering
\includegraphics[width=.85\textwidth]{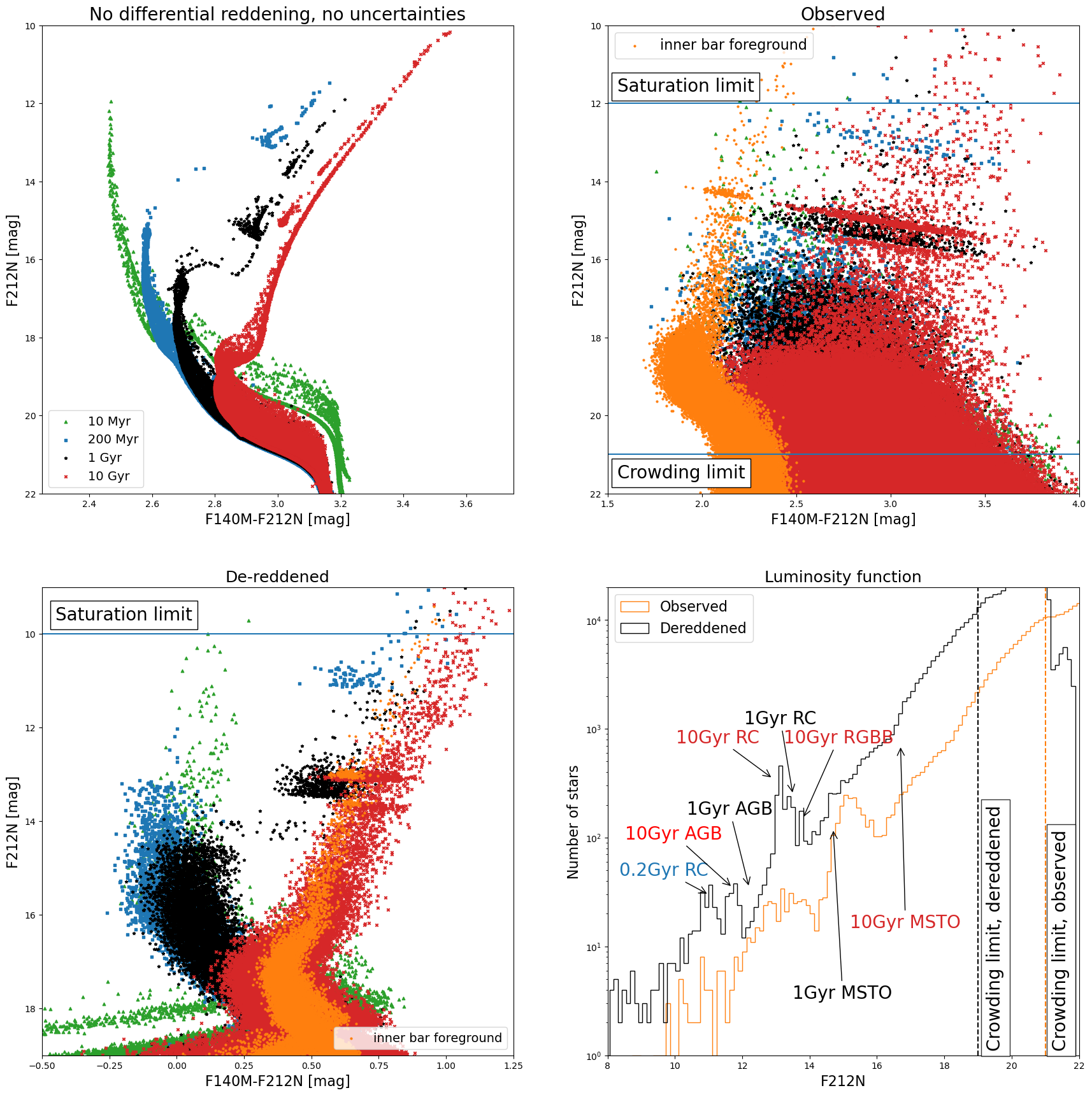}
\caption{Simulated CMDs and  luminosity functions. We used the SPISEA python package to simulate observations of the stellar population at the GC. We assumed a star formation history
  roughly similar to the one inferred  by \citet{Nogueras-Lara:2020pp} and \citet{Schodel:2023ov}, that is 1\% of the (originally formed) stellar mass forms at 10\,Myr, 4\% at 200\,Myr, 10\% at 1\,Gyr and 85\% at 10\,Gyr. We assumed a distance of $8.25$\,kpc, a mean extinction of $A_{K}=2$\,mag with Gaussian differential reddening  $\sigma A_{K}=0.2$\,mag \citep[estimates based on][]{Schodel:2014fk,Nogueras-Lara:2021di}, and constant photometric uncertainties of 0.01\,mag. We did not include any uncertainty of the distance modulus and assumed a perfectly-known and spatially non-variable extinction curve.
Upper left: CMD without differential reddening or photometric uncertainties.  Upper right: Observed CMD, after including differential reddening and observational uncertainties. Here, we have included a foreground population from the inner bulge, with the same properties as the 10\,Gyr NSD population, but at a distance of 8\,kpc, mean reddening of $A_{K}=1.3$\,mag, and with Gaussian differential reddening  $\sigma A_{K}=0.05$\,mag. These values have been derived empirically from GNS data \citep[see][]{Nogueras-Lara:2018pr,Nogueras-Lara:2022wa}. The latter work also demonstrates that we can reliably exclude additional polluting components from the Galactic foreground through color cuts.
Lower left: Dereddened CMD. The intrinsic color terms of almost all stars in the simulation are small and vary very little, as shown in Fig.\,\ref{fig:stellar_colors}. By assuming a constant color term ($[F212N - F480M]\approx-0.03$) for all stars, we can therefore  deredden the CMD.   Lower right: Observed and de-reddened luminosity function.  In the absence of accurate short-wavelength observations, we can infer the star formation history from the luminosity function \citep[for the methodology, see][]{Schodel:2018db,Nogueras-Lara:2020pp,Schodel:2020qc,Nogueras-Lara:2022jz}. There are several clearly visible markers of different star formation episodes (AGB, RC and RGBB bumps, MS turnoffs for different ages). Using the CMD can provide more information, but requires costly deep observations in the short NIR (F140M). \label{fig:obs}}
\end{figure}

Figure\,\ref{fig:obs} shows a simulation of a field observed at the GC with F140M, F212N and F480M. The observed CMD becomes severely blurred due to strong and differential extinction, and all stellar populations  overlap. Only after de-reddening can we arrive at an interpretable CMD and separate different stellar populations (at least partially).  The observed $[F212N-F480M]$ color is a very good tracer of reddening, because the intrinsic colors of all types of stars are small and almost constant with these filters (Fig.\,\ref{fig:stellar_colors}).

There exists an efficient method to infer the star formation history from a single epoch of JWST imaging: the use of the stellar luminosity function. This methodology has already been tested extensively for the GC with other instruments \citep[e.g.][]{Nogueras-Lara:2020pp,Schodel:2020qc}. We partially know and partially expect the star formation history to vary across the GC, from the immediate environment of Sgr\,A*, through the NSC, inner NSD, star-forming regions and regions containing young clusters, to the outer NSD. Therefore only a complete survey of the inner 100\,pc can provide an unambiguous picture. NIRCam is the only facility  with the  angular resolution and sensitive mid-infrared imaging capabilities required for accurate, reddening-corrected measurements of the stellar luminosity function in the GC (see Fig.\,\ref{fig:obs}). NIRCam will be able to detect the main sequence turn-off of the oldest population throughout the GC. Only a deep, accurately de-reddened luminosity function will allow us to identify different star formation epochs and show how the star formation history changes across the GC, that is, as a function of distance from Sagittarius\,A*: Does the NSD grow from the inside out?  

Proper motion measurements will provide us with a step change in understanding the star formation history because they will allow us to go beyond statistical corrections for the different components observed towards any given field and identify the kinematic sub-populations such as the bulge, NSD and NSC and study them separately \citep[e.g.][]{Shahzamanian:2022vz,Nogueras-Lara:2023xd}.

\subsection{How much stellar mass formed in the past $\sim$30\,Myr and what does this imply for the overall energetics of the GC?}
{\bf The NSD is the most prolific site of star formation in the Milky Way, but so far we lack a census of young stars in this region.} Observational evidence points to the existence of a significant number of clusters and associations with smaller masses or in a more advanced state of
dissolution than Arches and Quintuplet 
\citep[$\sim$$10^4$ M$_{\odot}$, e.g.][]{Dong:2011ff,Matsunaga:2011uq,Clark:2021fj,Nogueras-Lara:2022jz,Martinez-Arranz:2023wn}. Without NIRCam, they are undetectable, because they do not show up as
over-densities in the  GC, and it is barely possible to differentiate between massive young stars and older, cool giants in the GC photometrically, as the degeneracy between reddening and the stellar color is significant. While spectroscopy can discern the stellar types, it is impossible to cover the necessary large field with sufficiently sensitive spectroscopic observations at the required angular resolution of $\lesssim0.2"$.


NIRCam/JWST will allow us to identify hot, young stars:  (1) via accurate de-reddening (Fig.\,\ref{fig:obs}), combined with either ground-based \citep[GALACTICNUCLEUS][]{Nogueras-Lara:2019yj}, HST WFC3 F127M photometry   \citep[see, e.g., Fig.\,6 in][where the massive stars in the Quintuplet cluster can be traced down to the onset of the pre-main sequence in a $F153M$ vs. $F127M-F153M$ CMD]{Rui:2019ch}, or JWST NIRCam F140M photometry; (2) via precise proper motions from multi-epoch imaging, where we can identify clusters and associations as co-moving groups \citep{Martinez-Arranz:2023wn}.

Identifying these hidden young populations that trace star formation over the past few to tens of millions of years will allow us to understand whether star formation \citep[e.g.][]{Sarkar:2015oe} or (recurrent) black hole activity \citep[e.g.][]{Yang:2022jm} drive the outflow from the GC \citep[see][]{Heywood:2019mw,Ponti:2019ul,Predehl:2020py} and dominate the contemporary energetics of our Galaxy's nucleus, as well as to study the initial mass function across the GC region, as discussed in the next point.

\subsection{What is the origin and environmental variation of the stellar initial mass function?}

The stellar Initial Mass Function (IMF) describes the distribution of stellar masses that are created during star formation. 
Its properties are a reflection of the many physical processes involved in the star formation process, making it a critical observational benchmark for star formation theory \citep[e.g.][]{Krumholz:2014xn, Offner:2014rw}. 
In addition, the IMF is an underlying parameter used in many areas of astrophysics, such as the star formation history of the universe \citep[e.g.][]{Madau:2014fv}, galaxy mass assembly and evolution \citep[e.g.][]{Kauffmann:2003qr}, and compact object production and merger rates \citep[e.g.][]{Rodriguez:2016qq}. Thus, understanding the properties of the IMF and how it behaves in different environments has far-reaching implications for star formation theory and beyond. 

As the only galactic nucleus where we can resolve individual (forming) stars, the GC offers a unique opportunity to directly measure the IMF in an extreme environment with conditions more similar to those in high-redshift starburst galaxies than to star-forming regions near the Sun \citep{Kruijssen:2013tq, Henshaw:2022nm}.
Indeed, observations of the young stars in the central parsec and the two known young massive clusters in the CMZ suggest that they exhibit an overabundance of high-mass stars (or possibly a dearth of low-mass stars) compared to the ``standard'' IMF observed in the Milky Way Disk \citep[e.g.][]{Bartko:2010fk, Lu:2013fk, Hosek:2019vn, Gallego-Calvente:2021yl, Gallego-Calvente:2022al}.
These clusters are the subject of ongoing JWST programs. However, several key questions remain: how has tidal stripping impacted the star cluster populations, potentially biasing their IMF measurements? Do all star clusters near the GC show IMF variations, indicating that the environment is the cause, or is it simply a property of massive clusters?
The proposed survey addresses these questions by (1) allowing for the first detection and characterization of the tidal tails of the young massive clusters, which are expected to extend tens of parsecs from the clusters themselves \citep{Habibi:2014mb} and (2) measuring the IMF of previously-undetected lower mass young clusters in the region (e.g., $\mathsection$3.2) to determine if they also show IMF variations.

\subsection{Why is the star formation rate one to two orders of magnitude lower than predicted by standard star-formation-dense-gas relations?}
Our Galactic Center serves as a template for dense star-forming regions and galactic centers throughout the universe.  

Gas-star formation relations predict an order of magnitude higher star formation rate (SFR) than is presently observed, raising questions about the universality of these relations  \citep{Longmore:2013yu, Barnes:2017gl, Henshaw:2022nm}.
However, there is evidence that the SFR averaged over the last 30 to 40 Myr has been between 0.1 to 0.5 \SFR \citep{Matsunaga:2011uq,Nogueras-Lara:2020pp}, suggesting that time variation is part of the answer.
Yet the present-day SFR appears to be up to an order of magnitude lower.
JWST is needed to improve the precision of both the recent star formation history and  ongoing star formation to determine why the Galactic Center SFR is presently low. Greater precision will allow us to test whether time variability on scales of one to a few million years may play an important role.

Counting the number of YSOs will solve one of the longest-standing questions in Galactic Center star formation research: Why do standard SFR measurements based on IR luminosity, cm-continuum emission and high-mass YSO (HMYSO; $M_*\gtrsim$10 M$_\odot$) number counts all show the CMZ SFR is 1--2 orders of magnitude lower than predicted by dense gas star formation relations? As existing SFR measurements are completely dominated by the light from high-mass stars, they miss a population of low-mass YSOs (LMYSOs) left undetected by current facilities.  If JWST detects a large population of LMYSOs, the ‘depressed’ CMZ SFR conundrum may be resolved. However, such a solution implies that canonical SFR calibrations, upon which much of our understanding of galaxy evolution is based, underestimate SFRs by over an order of magnitude. If JWST confirms the previous SFR measurements and finds the expected population of LMYSOs extrapolating from known HMYSOs and extended star formation tracers, the SFR calibrations will be validated, but the mystery of GC SFR deepens. 

A fundamental prediction of star formation theories combining the scale-free physics of turbulence with gravity is an environmentally dependent volume density threshold for star formation. This ‘critical density’ depends on gas properties such as the mean density, virial ratio, and Mach number \citep{Federrath:2012nx}, so is predicted to be 10$^3$ times larger in the CMZ than the disk. Combining the JWST census of YSOs with measurements of the physical and kinematic gas properties from existing (sub)millimeter surveys, we will determine how the number and luminosity of YSOs vary with gas column density and velocity dispersion. Measurement of the relation between the YSO surface density and gas properties will determine the value of the environmentally dependent critical density.

It is extremely challenging to distinguish YSOs photometrically from highly reddened main sequence and giant stars.
This is an area of open discussion and the community still needs to identify the best strategy to distinguish these populations. 
Nevertheless, JWST is poised to play a critical role in changing our view of star formation in the Galactic center by enabling star-counting based measurement of the star formation rate and detailed study of how and where stars of different masses form.
The current state of the art using Spitzer is limited to the most intrinsically luminous sources \citep{Yusef-Zadeh:2009fk} and is subject to substantial contamination \citep{Koepferl:2015aq}.
Spectroscopic observations have shown that the star formation rates estimated from Spitzer observations are entirely dominated by contaminants \citep{An:2017gt,Jang:2022jw}, but JWST's high resolution, especially at 4.8$\mu$m, will enable us to separate genuine massive YSOs from background sources shining through clouds.
JWST should readily detect the vast majority of $L>1$ L$_\odot$ YSOs with the planned shallow exposures and is capable of detecting YSOs down to $\sim0.1$ L$_\odot$ in many cases (see Appendix \ref{sec:ysoid}).
Further study is needed, however, to determine the most efficient NIRCam filter combinations to distinguish these from deeply embedded post-main-sequence stars.

\subsection{What is the 3D structure of the interstellar medium (ISM) orbiting and fueling accretion and star formation at the Galactic Center?}
Many of the open questions about GC star formation stem from uncertainty in the relative position of gas and stars and their proximity to the center.
The strength of different physical mechanisms, including gravity and radiation pressure, and the effectiveness of feedback from the central black hole versus supernovae, all depend on the precise locations of gas clouds as they traverse the nuclear disk \citep[e.g.][]{Kruijssen:2015fk}.

Mapping the 3D gas distribution in the central few hundred parsecs of the Galaxy reveals the interplay between gas, stars, and the central black hole, offering insights into the GC's structure, dynamics, and formation history \citep{Henshaw:2022nm}. The gas distribution unveils the formation and evolution of key structures like molecular clouds and dust lanes, fueling gas accretion and star formation. Furthermore, studying the structure and dynamics of the interstellar medium in this region provides an understanding of mass distribution, gravitational potential, and dark matter content, contributing to the broader understanding of galaxy formation and evolution.

The previously discussed JWST NIRCam star counts combined with measurements of reddening to individual stars will lead to a measurement of the relative position of clouds along the line of sight: clouds located in  front of the NSD will have relatively few stars with low reddening and a large number of stars with high reddening while clouds located in or behind the NSD will have a large number of stars with low reddening.  Given the high absolute stellar density, it will be possible to make 3D maps of cloud location and structure. Proper motions can provide additional constraints on the position of molecular clouds along the line-of-sight \citep{Martinez-Arranz:2023wn}.

The 3D distribution of molecular clouds can, on the one hand,  be used to interpret the variable signal from X-ray reflecting clouds \citep[e.g.][]{Ponti:2013hg} and must, on the other hand, be consistent with time variability measurements and polarization measurements (IXPE) in the X-ray domain \citep{Marin:2023mj,Khabibullin:2025lk}. The 3D distribution of molecular clouds
can  be used to see how they constrain and redirect outflows from the innermost regions.

NIRCam astrometry, combined with $0.2"$ astrometry from the ground (GALACTICNUCLEUS, see section\,\ref{sec:GNS}) will 
provide stellar proper motions with a precision of better than $\sim$10 \kms\  (0.25\,mas\,yr$^{-1}$).
A second epoch of JWST imaging in at least one band will be needed to measure proper motions in the densest parts of clouds, where GALACTICNUCLEUS detects no stars. 
Combining the JWST proper motions with ALMA radial velocities of clouds, which have relative line-of-sight distances measured, will lead to a determination of the location of the cloud relative to the nucleus as well as provide constraints on the cloud's 3D space velocity.   The identification of young stellar objects (YSOs) embedded in the cloud, combined with the proper motions of these YSOs and the cloud's radial velocity will lead to a measurement of the cloud's 3D velocity vector. 

The combination of JWST reddening, star counts, and proper motions, combined with ALMA radial velocities will enable the production of a face-on view of the CMZ clouds along with measurement of the 3D motions of these clouds.  For the first time, we will be able to measure all phase-space dimensions of the CMZ dense gas, enabling a direct comparison with models of gas flows in the Galaxy's barred potential. 

\subsection{What does the GC teach us about extragalactic astrophysics?}
As the only center of a galaxy that can be resolved into individual stars and studied on scales of milli-parsecs, the GC is an indispensable template for the astrophysics of galactic nuclei (at the very least for those galaxies that are similar to the Milky Way). The science cases discussed in this White Paper are not only intrinsically important by themselves, but they all contribute together to our understanding of fundamental open astrophysics questions: {\it How does a galactic nucleus work? In what way are astrophysical processes different in extreme environments?} NIRCam/JWST can provide us with the most detailed insights into the structure and kinematics of a galaxy nucleus to probe its precise gravitational potential and to see how the individual components form and interact with each other (Central Molecular Zone, young  massive clusters, nuclear star cluster, nuclear disk, central black hole). With accurate measurements of the amounts of stellar mass formed in the past few 10 Myr, we will be able to constrain whether it is star formation or sporadic black hole activity that dominates the energetics of quiescent nuclei -- by far the most common nuclei in the present-day Universe. Understanding how extreme conditions of the ISM translate into outcomes of star formation, such as its efficiency and the IMF, will allow us to learn basic lessons from the GC laboratory that can be applied to galactic nuclei in the nearby Universe and even to star formation in the early Universe. Having a detailed understanding of the physics of the Central Molecular Zone from our GC is necessary to understand the CMZs of other galaxies and their similarities and differences with our own. For example, the nearby galaxy NGC\,253 is often considered a Milky Way analogue. Its CMZ is very similar to the one of our Galaxy, but unlike our GC, the present-day star formation  rate in the nucleus of NGC\,253 is more than ten times higher. Is this offset the result of a consistently elevated star formation efficiency in NGC 253 compared to the Milky Way's GC? Or are we witnessing the two nuclei in different stages of a similar cycle of SF variabilty from quiescent (Milky Way) to starburst (NGC 253)? The proposed survey will therefore provide us with insights that will be of general interest to astrophysics.

\subsection{Other questions.} Beyond the major questions mentioned above, the proposed survey will allow the community to address numerous other science cases. Here we briefly mention some of them: (1) How does stellar feedback affect the evolution of the ISM structure in the GC? 
(2) What is the origin of the enigmatic non-thermal filaments: The inner few hundred light years of the Galaxy hosts hundreds of mysterious magnetized radio filaments \citep{Heywood:2019mw}; their intrinsic polarization shows that their magnetic fields are directed along the filaments. In one proposed scenario,  the interaction of a cosmic-ray-driven wind with stellar wind bubbles creates magnetized cometary tails \citep{Yusef-Zadeh:2019uy}. NIRCam images will  examine this scenario by identifying  which mass-losing stars are associated with compact radio sources and filaments. (3) How does the extinction curve vary across the GC and what does this tell us about the variation of dust properties as a function of environment within galaxies? The extinction curve towards the GC is still uncertain. Narrow and medium band mid-infrared photometry with NIRCam will greatly help to constrain the wavelength dependence of reddening in this region \citep[e.g.][]{Nishiyama:2009oj,Nogueras-Lara:2020qn}.


\section{Setup of JWST Galactic Center Treasury Survey}

\subsection {Survey area} 

\begin{figure}[htb]
\includegraphics[width=\textwidth]{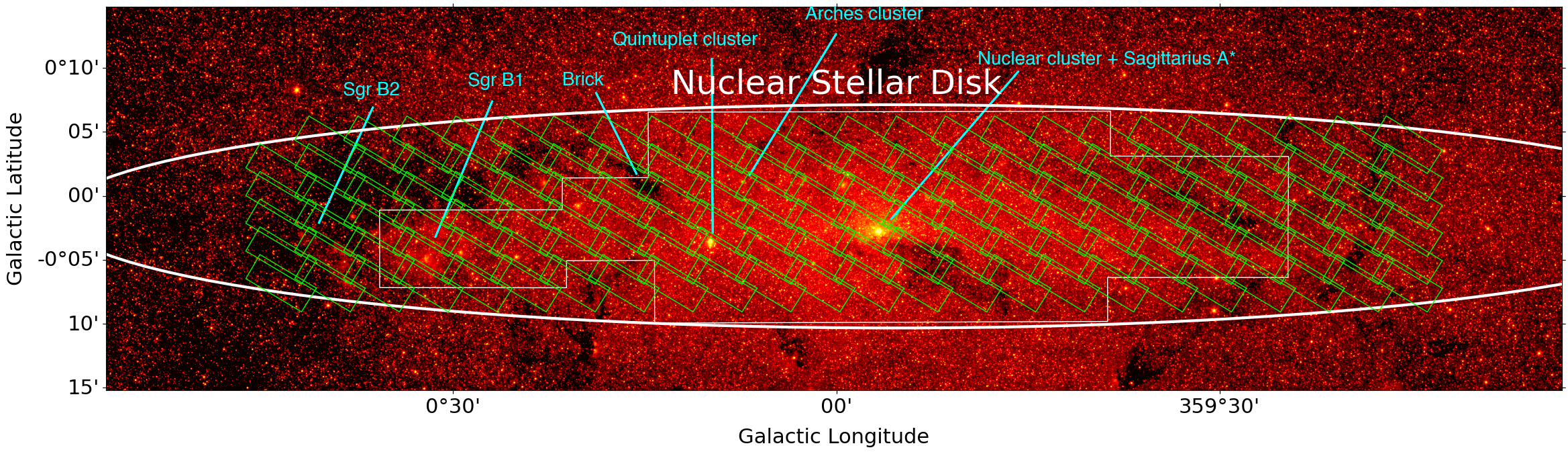}
\includegraphics[width=\textwidth]{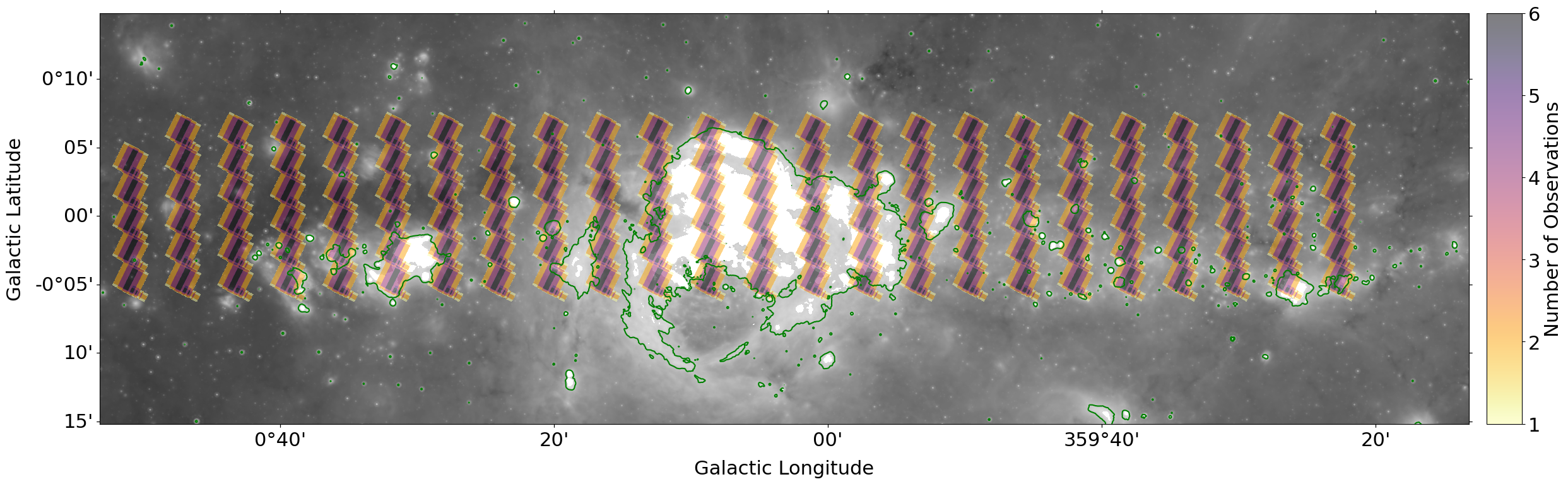}
  \caption{\label{fig:obsoverview}{\bf Overview of the proposed target fields.}
  Upper panel: Proposed NIRCam pointings superposed on a Spitzer IRAC $4.5\,\mathrm{\mu}$m image of the GC. The white polygon indicates the area of the GALACTICNUCLEUS survey \citep{Nogueras-Lara:2019yj}. The green
    rectangles indicate the NIRCam target fields, assuming a FULLBOX 6TIGHT
    dither pattern, which we consider to provide the best compromise
    between sampling and speed. Lower panel: Coverage map of parallel MIRI pointings superposed on a Spitzer MPIS $24\,\mathrm{\mu}$m image.}
\end{figure}

The properties of the stellar populations in the GC depend on position (intrinsic changes and different projected mixtures of inner bar, NSD and NSC populations) and so do the properties of the ISM (e.g.\ dependency on distance from \sgra). The different molecular clouds are in different phases of star formation, the young clusters and associations are sparsely distributed, extinction is patchy and varies strongly across the field and de-projection of stellar structures requires a knowledge as complete as possible of the GC.
 Therefore addressing our community's science questions requires a NIRCam survey of the entire Galactic Center out
to the edge of the 100\,pc stream of the CMZ. This corresponds to an area of about $1.25^{\circ}\times0.25^{\circ}$ (Fig\,\ref{fig:obsoverview}), or about 180\,pc$\times$36\,pc.

\subsection {Filters}

The completeness of the proposed observations will be limited by stellar confusion. Since the target is also comparatively bright, narrow and medium-band filters will provide sufficiently high signal-to-noise for all objectives while targeting key diagnostic lines of the ISM and key sections of the stars' continuum radiation. These filters have the added advantage of well-defined effective wavelengths (which are a function of both stellar type and reddening) and minimised saturation. With the chosen filters we will (hard-)saturate stars brighter than $F212N\approx11$\,mag.
Preliminary work with F212N NIRCam observations of Sgr\,C have taught us that we can repair the PSFs of at least two magnitude brighter saturated stars. This is both important for obtaining measurements of these stars and, even more so,  for removing the bright PSF features around these stars, which hamper the detection and measurement of fainter stars. Stars brighter than $F212N=11$\,mag
make up $<<0.1\%$ of all stars that we expect to detect in the survey area, assuming the star formation history of \citet{Nogueras-Lara:2020pp}.  The small number and surface density of such bright stars implies that saturation and extended bright star PSFs will have a minimal impact on our science goals.

Due to the large survey area, sub-pixel dithering should be avoided, because it would increase the required observing time by factors of a few. For optimal astrometry and photometry, the filters should therefore be chosen at wavelengths longer than (or close to) the Nyquist wavelengths of the NIRCam NIR and MIR channels, at $2$ and $4\,\mu$m. 

We propose to carry out the survey with four or five filters: F140M, F187N, F212N, F405N and F480M. Sufficiently deep observations at the shortest wavelength (F140M) may possibly be obtained with HST WFC3 and the F127M filter. Observations with the NIR AND MIR filters will be carried out simultaneously, taking advantage of  NIRCams two modules.  F212N and F480M are excellent tracers of the stellar continuum. Tests with atmosphere models using the SPISEA code \citep{Hosek:2020mi}, show that the magnitude difference between the medium band filter F210M and the narrow band filter F212N is $<0.05$\,mag for the stars that can be detected by NIRCam at the GC. Even though F212N is centred on an H$_{2}$ line, there is little such emission at the GC, as existing NIRCam observations of the GC show (Hosek M. W. Jr., Ginsburg A.,priv.\ comm.). Also, PSF fitting tools such as StarFinder will take any diffuse emission underlying point sources into account. Since the intrinsic color at F212N-F480M of practically all stars brighter than the crowding limit is almost negligible and practically constant (Fig.\,\ref{fig:obs}), these filters will allow us to measure the reddening of each star individually. F187N and F405N are centered on the Paschen-$\alpha$ and Brackett-$\alpha$ H\,I lines and will allow us to infer the conditions of the ISM (together with the F212N filter, which traces rovibrational emission of \Htwo\ in shocks and photo dominated regions),  and find massive post main sequence stars via the line emission in their stellar winds \citep{Dong:2011ff}, and possibly identify YSOs from excess emission from accretion \citep[e.g.][]{Alcala:2017yc}.

\subsection{Photometry}

With our chosen setup (see below)  we can reach SNR $\approx100$ for stars of $F212N=18$\,mag and $F480M=18$\,mag, corresponding to a photometric uncertainty of the order $0.01$\,mag. With these values, we can
estimate {\it relative} reddening towards these sources at all near-infrared bands  with uncertainties
$\lesssim0.05$\,mag. The systematic uncertainties of interstellar extinction depend on the conversion factor from color to extinction and on the uncertainties of the extinction curve towards the GC.  They are on the order of 10--15\% for the considered wavelengths, or about $0.2$ to $0.3$\,mag for an extinction of $A_{2\,{\mu}m}=2$\,mag.

The survey will fully cover red clump stars in all four requested filters, giving us a unique opportunity to characterise the extinction curve in the NIR and mid-infrared to solve the apparent disagreement between the many extinction laws in the literature at these wavelengths \citep[e.g.][]{Nogueras-Lara:2020qn}.

 The confusion limit of NIRCam/JWST in the
targeted area is $F212N=20-21$\,mag \citep[inferred from JWST observations of Sgr\,C, priv.comm., and extrapolation from][]{Nogueras-Lara:2019yj,Schodel:2020qc}, which means we will observe stars as faint as late G-type on the main sequence.  We will therefore be able to use the
luminosity function to constrain the presence of a main sequence
turn-off as old as $10$\,Gyr (Fig.\,\ref{fig:obs}).

\subsection {Astrometry}

Astrometry will be done in F212N (except for highly embedded sources, which may be significantly brighter in F480M). With an angular resolution of $0.07"$, we can thus measure relative astrometric positions with a precision of $<1$\,mas for all sources with a signal-to-noise $>100$, corresponding to an observed F212N magnitude of about $18$\,mag, where we expect to be well above the confusion limit. Relative precision alignment between different epochs and instruments can be obtained with several thousands of stars per image from each
epoch. With relative astrometric uncertainties of $\lesssim1$\,mas
for the stars in two given epochs, the alignment uncertainty can then
be kept below 1\,mas \citep[considerations based
on][]{Schodel:2009zr,Shahzamanian:2019cy,Shahzamanian:2022vz}.  With two epochs spaced five (ten) years apart, we can thus obtain proper motion measurements with an accuracy of better than $0.3$(0.15)\,mas\,yr$^{-1}$ (conservative estimation based on the faintest stars) for an estimated ten million stars (extrapolated from GALACTICNUCLEUS, see section\,\ref{sec:GNS}). This precision is well below the velocity dispersion of the NSD and inner bar and will allow us to find young star clusters and associations in the form of co-moving groups. We will also be able to measure the proper motions of deeply embedded YSOs and thus of the associated molecular clouds.

\subsection{ Strategy \label{sec:strategy}}

\begin{SCfigure}[0.5][htb]
  \centering
\includegraphics[width=.5\textwidth]{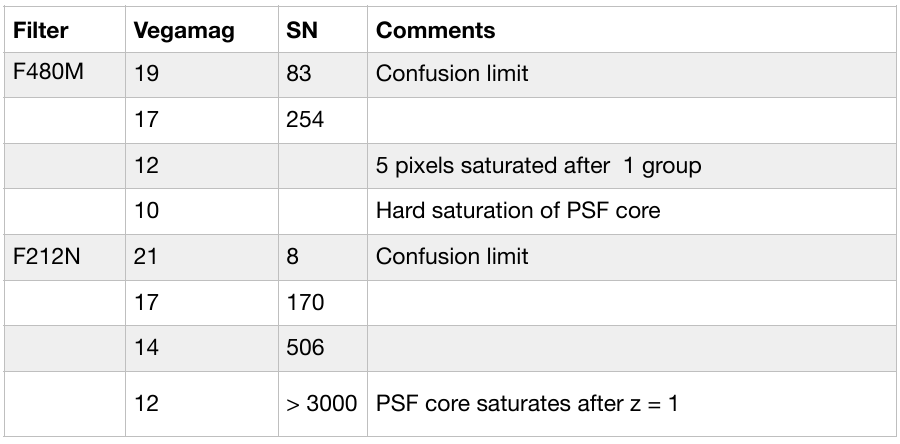}
  \caption{\label{Tab:ETC}Filters, magnitudes and SNR for the chosen observational setup of epoch\,1 (BRIGHT2, 4 groups per exposure, 2 exposures per dither, 2 dithers).}
\end{SCfigure}

Since the area of interest is significantly larger than the NIRCam
field of view, we have to cover the whole region with a mosaic of
independent pointings. The FULLBOX 6TIGHT pattern with no sub-pixel dithering appears to provide the best
trade-off between efficiency,  SNR, suppression of artifacts and homogeneous depth. The different pointings are
chosen to overlap in right ascension and
declination, which will serve to obtain a homogeneous coverage of the entire area (the edges of a field observed with FULLBOX 6TIGHT are covered by less dithers than the more central parts).  We choose the BRIGHT2 readout pattern with 4 groups per integration to minimize the saturation of bright stars, while at the same time allowing us to use parallel MIRI observations (not possible with BRIGHT1 due to the high data rate). Our setup means that almost the entire field is covered by at least four dithers (see \S\ref{sec:mosaic}). With 2 integrations per dither we can thus reach a minimal signal-to-noise ratio of 
$8$ at  $F212N=20$\,mag and $83$ at $F480M=19$\,mag, at which magnitudes we will reach the confusion limit. Table\,\ref{Tab:ETC} provides an overview of exposure times and signal-to-noise.

The proposed GC survey should be carried out over various epochs to enable proper motion measurements, to decrease risk, and to facilitate the allocation of observations. Since practically all science cases require an accurate determination of interstellar extinction/reddening we consider imaging at F212N and F480M to have the highest priority and it should be carried out in the first epoch.
The observations at the shortest wavelengths, where extinction is highest and source confusion therefore lowest, could potentially be carried out with WFC3/HST at F127M.

A second epoch will be observed using F187N+F405N as a default option. These filters are sensitive to extended recombination line emission, and their ratio in HII regions is fixed, enabling extended extinction mapping. 
This second epoch will use a very similar instrumental setup, but we leave open the possibility that further studies will identify a different set of filters better optimized to completing the science goals above.

A third and, possibly, fourth epoch will be observed five and ten years after the first epoch in an identical way to the first one. These epochs will serve for proper motion measurements. While epoch 2 can be used, in a limited way, for proper motions, it may be too close in time and with different filters and therefore not ideal for proper motion measurements. One of the epochs may include the F140M filter instead of F212N, to obtain the best possible wavelength leverage for CMDs.

According to our preferred APT setup, each epoch will require about 137.5\,h hours of charged time with about 41.5\,h of science time.


\section{Synergy with existing VLT, HST, and Roman surveys, proper motions}
\label{sec:GNS}

The proposed NIRCam survey will profit greatly from synergies with existing data from ground and space based observations, in particular the following ones:

\begin{itemize}
    \item The HST Paschen-$\alpha$ survey \citep{Wang:2010fk,Dong:2011ff} covers an area of $\sim$$39\times15$\,arcmin$^{2}$ ($90\times$35\,pc$^{2}$) with NICMOS F187N and F190N observations. Its catalog contains $\sim$$570,000$ point sources.
    \item \citet{Libralato:2021qa} present a catalog of $\sim$$830,000$ point sources in two epochs from HST WFC3 observations, covering three disconnected fields of $\sim$$10\times5$\,arcmin$^{2}$ and one of $\sim$$5\times5$\,arcmin$^{2}$ at the GC down to magnitudes of $[F139M]$$\sim$22\,mag. They measure the proper motions of about $120,000$ stars with an uncertainty of $\leq0.7$\,mas\,yr$^{-1}$.
    \item The GALACTICNUCLEUS survey \citep{Nogueras-Lara:2019yj} covers an area of $\sim$0.3\,deg$^{2}$, or about 6000\,pc$^{2}$. It provides photometry in the J, H, and Ks bands with photometric uncertainties better than 0.1\,mag (J) and $0.05$\,mag (H and K$_{s}$) for over 3 million stars. At $0.2"$ angular resolution, GALACTICNUCLEUS is the most complete survey of the GC up to date. A second epoch of H-band GALACTICNUCLEUS observations was acquired in 2021/2022 for proper motion measurements (PI Shahzamanian, ESO ID 105.20CM.001, not published yet).
    \item The ``Near Infrared Inner 600 pc Photometric Legacy Survey'' (PI M.~Zoccali, ESO Programme ID 0103.B-0262) used the GRAAL ground layer AO system of the wide field NIR camera HAWK-I at the VLT to image the central $4\times0.5$\,deg$^{2}$ of the Milky Way at angular resolutions of $0.3-0.4"$ in the H and Ks bands. The data have not yet been fully published.
\end{itemize}

The data from these surveys can be combined with NIRCam imaging to provide multi-band, multi-epoch photometry and multi-epoch astrometry. This will boost  the value of the NIRCam data considerably by improving the classification of stars and enabling proper motion measurements. Proper motion measurements, in particular, will provide a great added value to the first epoch of NIRCam imaging, because they will allow us to disentangle different stellar populations, infer the 3D structure of stars and molecular gas, and identify co-moving groups of young stars.

 GALACTICNUCLEUS covers about 70\% of the field proposed for the JWST NIRCam survey.  There are two significant synergies between GALACTICNUCLEUS and the proposed  NIRCam observations. (1) GALACTICNUCLEUS provides photometry at wavelengths shorter than $2\,\mu$m. In combination with the de-reddening of individual stars provided by the proposed F212N and F480M photometry, this will mean improved stellar classification (intrinsic stellar color differences for different types of stars reach up to $\sim$0.7\,mag in $[J-Ks]$, see Fig.\,\ref{fig:obs} ). (2) There exist already two epochs of GALACTICNUCLEUS (2015, 2021). Combined with a single epoch of NIRCam imaging, we can obtain precision proper motions  with an uncertainty $\lesssim0.3$\,mas\,yr$^{-1}$ for about $1\times10^{6}$ stars, as described below.

The {\bf Roman Space Telescope} \href{https://roman.gsfc.nasa.gov/science/galactic_plane_survey_definition.html}{Galactic Plane Survey} will image the GC with the F129, F158, F184, and F213 filters (R. Benjamin, priv.comm.). The GC NIRCam observations proposed in this White Paper will provide the community with essential complementary data to break degeneracies between stellar colours and extinction in these bands (as illustrated in Fig.\,\ref{fig:obs}; changing NIRCam F212N/F140M to WFI F213/129 will result in no change in the CMDs) and to identify closely blended stars in the lower resolution Roman images.

{\bf Proper motions}. The velocity dispersion of Bulge stars perpendicular to the Galactic Plane (and thus to the NSD) is
$\sigma_{\perp,GP,Bulge}\approx3$\,mas/yr. The velocity dispersion of the NSD is significantly smaller
$\sigma_{\perp,GP,Bulge}\approx1.5$\,mas/yr
\citep{Shahzamanian:2022vz}. The internal velocity dispersion of clusters is $\lesssim0.2$\,mas\,yr \citep{Rui:2019ch,Hosek:2019vn}. In order to be able to separate stellar populations and identify co-moving groups, we therefore aim at a proper motion uncertainty of better than $\sim$$0.3$\,mas\,yr$^{-1}$ \citep[see also][, who argue for a similar precision]{Matsunaga:2018dk}.

We discuss proper motion measurements using the example of GALACTICNUCLEUS here. GALACTICNUCLEUS is ideal for proper motion measurements, because of its large area and high angular resolution, and because the data reduction pipeline minimises residual differential atmospheric tip-tilt effects. Relative astrometric uncertainties of $\lesssim 1 (2)$\,mas are obtained in GALACTICNUCLEUS for stars of $H\lesssim 18 (19)$\,mag \citep[$H$ is the preferred band for astrometry in GALACTICNUCLEUS, see][]{Shahzamanian:2022vz,Martinez-Arranz:2022wj}. 
Assuming a mean extinction of $A_{Ks}=2$\,mag, $H\lesssim 18 (19)$\,mag  corresponds to [F212N]$\lesssim$16.5 (17.5)\,mag, fainter than the Red Clump at the distance and extinction of the GC. \citet{Shahzamanian:2022vz} presented a pilot proper motion study for the area common to GALACTICNUCLEUS and the HST Paschen-$\alpha$ survey. Major limitations were the relatively low SNR and small field-of-view of the NICMOS data. In particular the latter increased the uncertainty of epoch alignment. When combining NIRCam observations with GALACTICNUCLEUS, these problems will not be important. Due to the sensitivity and large field-of-view of NIRCam, we will be able to use on the order one thousand star for precise alignment between the data sets, thus limiting alignment uncertainties to $\lesssim1$\,mas.

Assuming that astrometric uncertainty scales as the FWHM of the PSF divided by the SNR \citep[e.g.][]{Fritz:2010fk},  we can reach a relative astrometric precision better than 1\,mas with the chosen filter and detector setup (see table \ref{Tab:ETC}).
The  GALACTICNCULEUS data are from  2015/2016. Assuming a NIRCam/JWST epoch of 2025 and measurement precision in
NIRCam and HAWK-I/VLT data of $\lesssim1$\,mas plus astrometric
alignment uncertainties of about 1\,mas/yr, we will therefore be able to measure proper motions to $\sqrt(3\times1^{2})/10.\approx0.2 (\sqrt(2\times1^{2} + 2^{2})/10.\approx0.25)$\,mas/yr for the $8\times10^{5} (1.1\times10^{6})$  GALACTICNUCLEUS stars brighter than $H=18 (19)$\,mag. These include all giants in the Red Clump and brighter. 

With estimates based on Sgr\,C observations, the survey area, and the SFH, we arrive at a number of 10 million stars for which proper motions could be measured with a second epoch of NIRCam imaging. The uncertainties of stars above the confusion limit ($[F212N]\geq20-21$) would be $\lesssim0.35$\,mas\,yr$^{-1}$, assuming 1\,mas relative astrometry, 1\,mas alignment uncertainty etween epochs and a five year time baseline. Such revolutionary data would enable exploring the GC kinematically down to almost solar mass main sequence stars.

\subsection{Data analysis, Techniques, and  High-Level Data Products}

This project will create and disseminate high level data, specialised tools and techniques as well as astrophysical  results applicable beyond its immediate objectives, such as:

\begin{itemize}
    \item Calibrated images for all filters and fields
    \item Astrometric and photometric catalogues cross-referenced with other existing catalogues
    \item Tools and techniques for crowded field photometry (the Galactic Center is possibly the most crowded field that can be observed)
    \item Tools to repair, fit, and subtract saturated stars
    \item Characterisation of the infrared extinction curve
    \item Individual star extinction measurements, detailed extinction maps
\end{itemize}

\subsection{A community resource}

The project proposed in this White Paper has implications for the entire GC community - and beyond. The collaboration will be open for other researchers to join. This approach to community engagement will be modelled after similar efforts in, e.g., the ALMA community. The fundamental philosophy in these teams is that every person who is willing to contribute in substantial ways can join working groups and the resulting publications.  Community outreach will chiefly be initiated via a website that will contain all relevant information, such as members of the collaboration and their contact, status of data acquisition/reduction/analysis, and publications. A prominent feature of this website will be a contact form for researchers interested in joining. Our openness to community engagement will also be featured prominently in the presentations we will be giving about this project.

\bibliography{BibGC}

\clearpage
\appendix




\section{Appendix: Detection \& Classification of YSOs}
\label{sec:ysoid}

We consider both the detectability and classification of YSOs in this Appendix.

To determine roughly what the luminosity detection limit is of YSOs, we use the \citet{Robitaille:2017lf} model grid.
This model grid includes a broad swath of all possible or plausible YSOs with disks and envelopes; it does not incorporate any information about protostellar evolutionary tracks and therefore includes some models that are physically unlikely.
The grid is also biased in number toward more highly embedded and extincted sources.
Nevertheless, it gives a conservative first estimate of what we may expect to see in the CMZ.

\begin{figure}[hp]
    \includegraphics[width=0.49\textwidth]{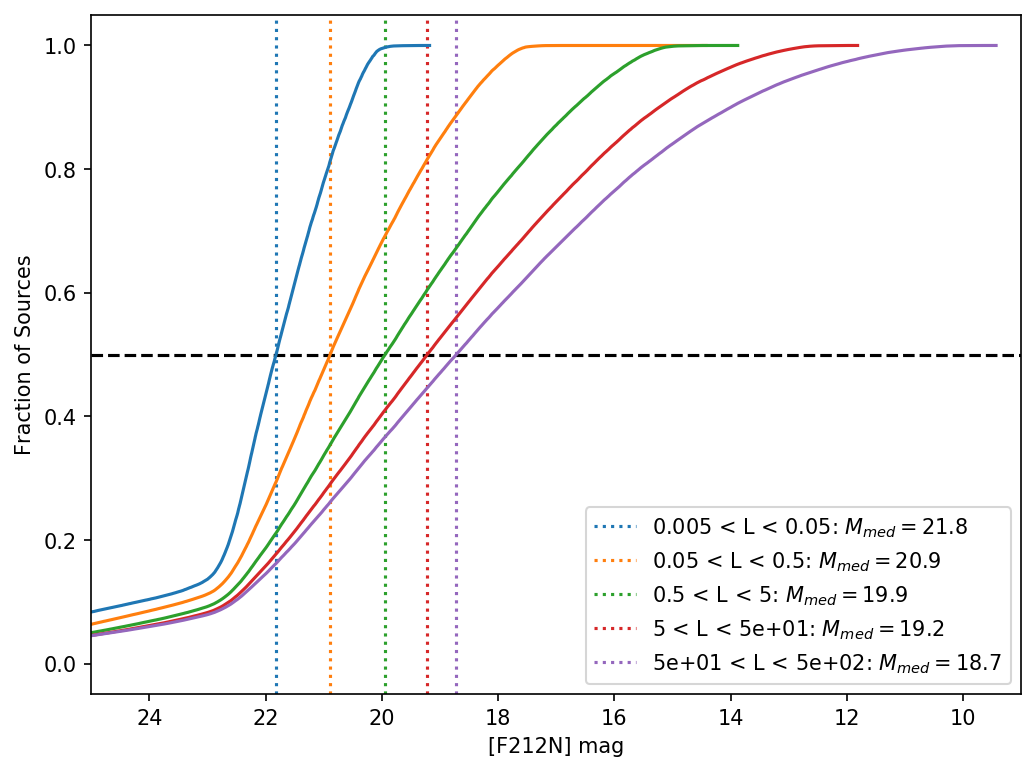}
    \includegraphics[width=0.49\textwidth]{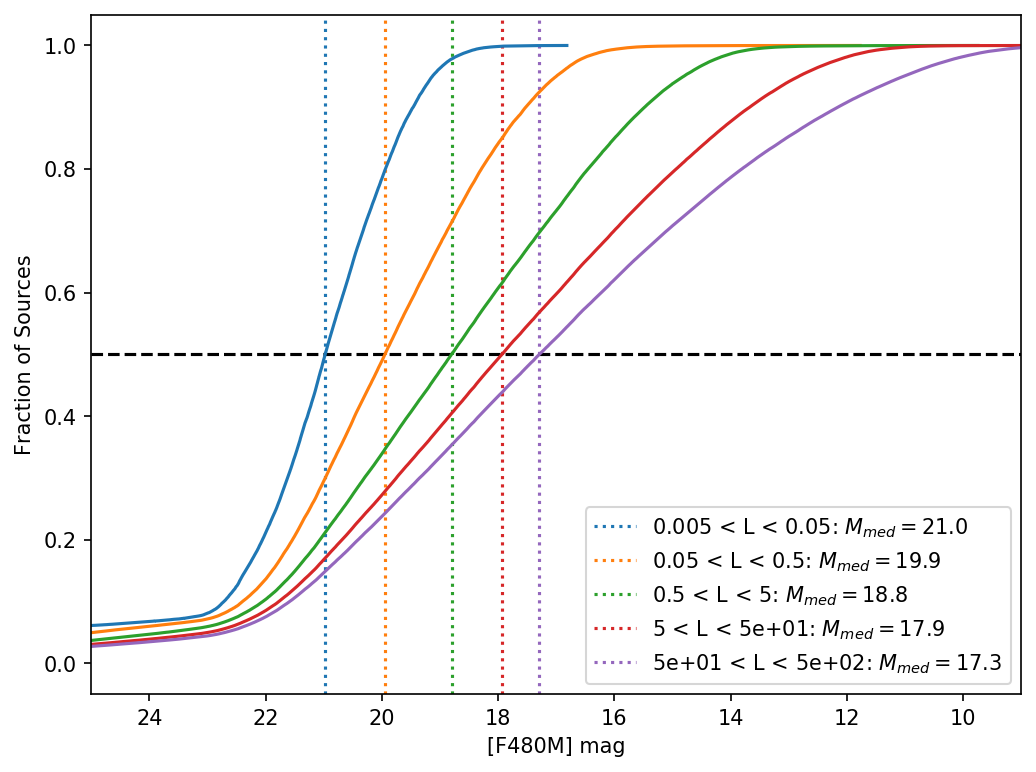}
    \label{fig:ysodetectability}
    \caption{Cumulative distribution functions of the observed magnitudes of YSOs at the distance of the CMZ based on the \citet{Robitaille:2017lf} model grid.
    Each curve represents a luminosity bin as labeled in the legend.
    }
\end{figure}

Figure \ref{fig:ysodetectability} shows the expected detection fraction of YSOs of varying luminosities.
For all central star luminosities, there is some fraction (roughly 10\%) that is not detectable by JWST at all: these are primarily edge-on disks and/or extremely thick envelopes.
The models shown do not include interstellar extinction, which should be roughly $A_{F212N}\sim3.5$ and $A_{F480M}\sim1.5$ for an assumed $A_V=30$.
A useful fraction - likely $>50\%$ - of YSOs will be detectable down to the confusion limit of 20 magnitudes for YSOs with $L\gtrsim1 \mathrm{L}_\odot$.
In F480M, the fraction is similar because of the higher confusion limit and lower extinction.

\begin{figure}
    \centering
    \includegraphics[width=0.32\textwidth]{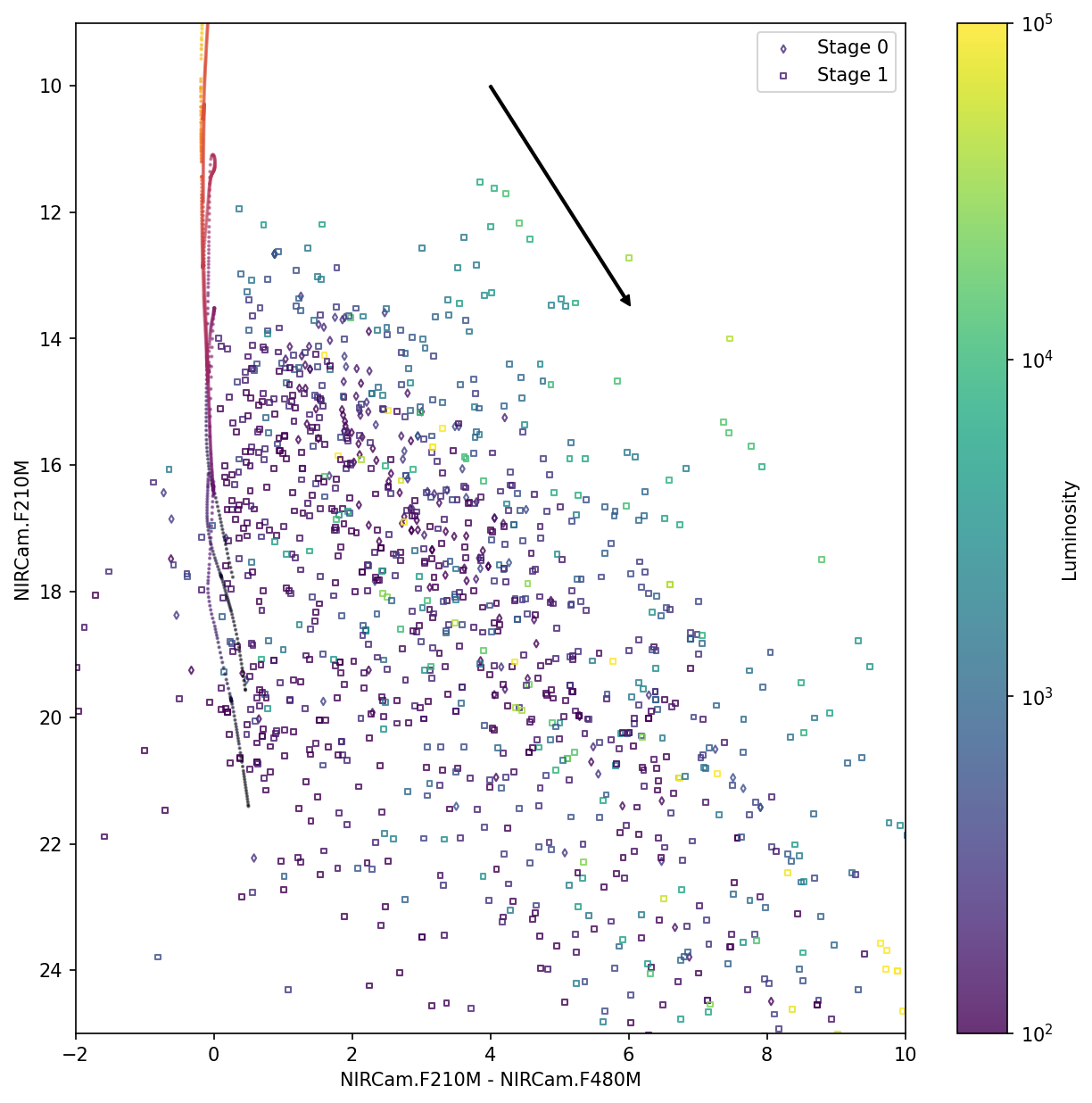}
    \includegraphics[width=0.32\textwidth]{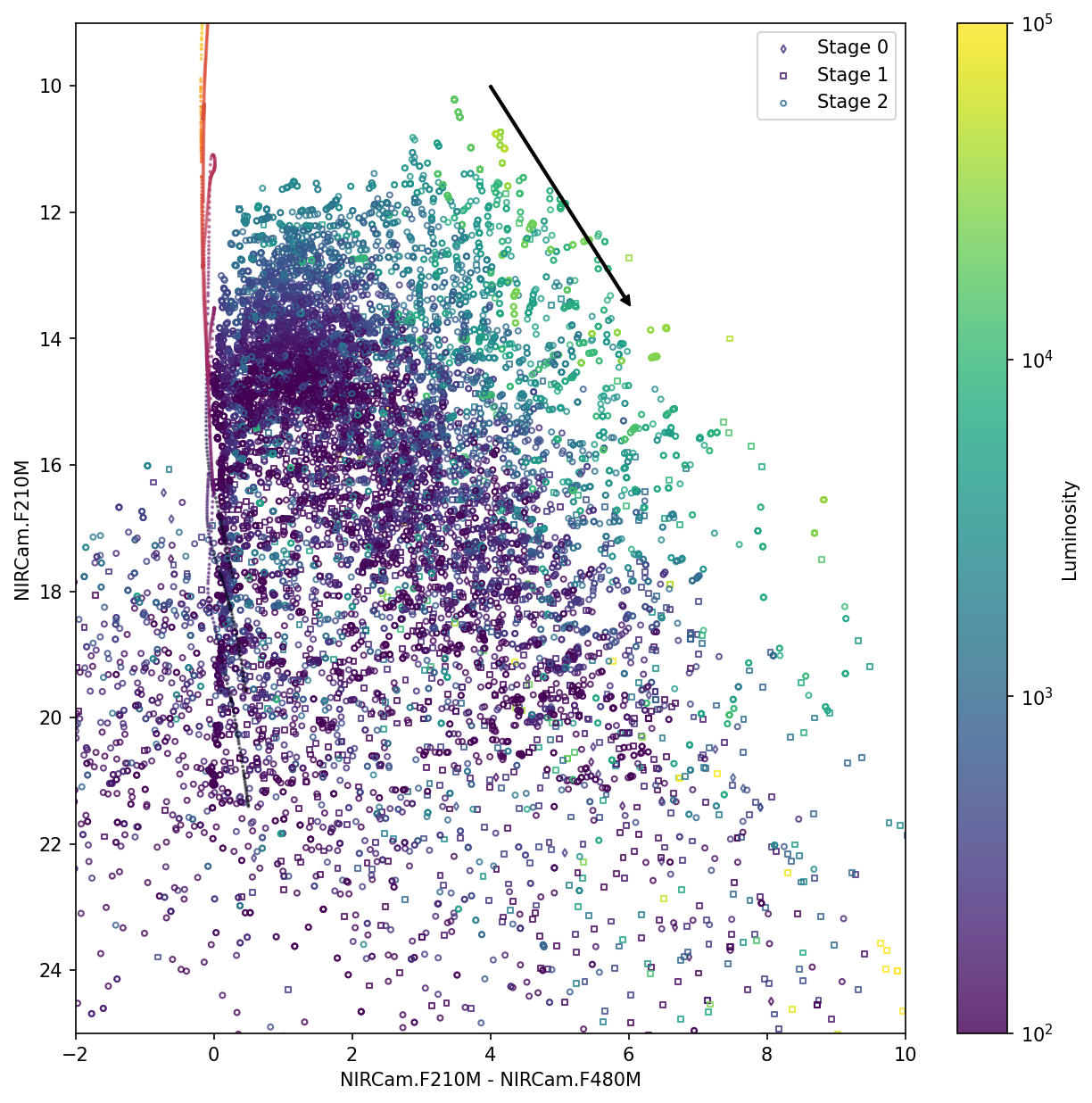}
    \includegraphics[width=0.32\textwidth]{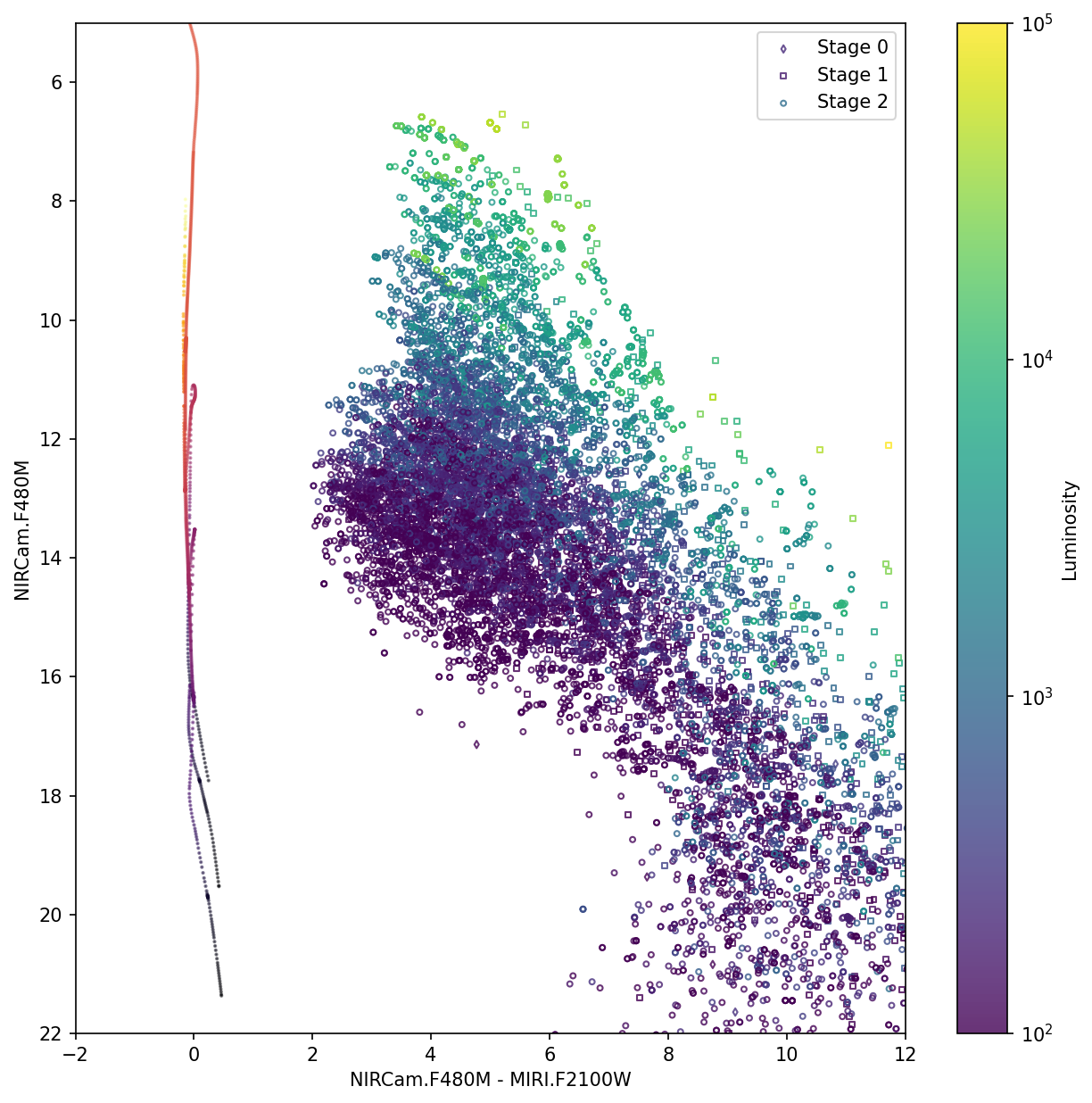}
    \caption{ Color-magnitude diagrams showing a synthetic population of YSOs in the selected NIRCam and MIRI filters.
    Left is Stage 0+I, middle is Stage 0+I+II, and right is all stages including the MIRI F2100W filter.
    Only YSOs with $M>4$ M$_\odot$ are included.
    This selection is intended to coarsely represent the expected population across the whole CMZ.
    The stage 0 and I sources are likely to be selectively more embedded, and therefore their detectable population may be smaller.
    The stage II sources are more likely to be distributed around the whole central hundred parsecs, as they may have had time to scatter from their birth environments.
    The black arrow in the left two plots is an $A_V=30$\,mag extinction vector using the CT06 extinction curve.
    The orange-to-black curves on the left side are MIST isochrones at $10^5, 10^6, 10^7$ yr.
    This figure shows that YSOs will be \emph{detectable}; classifying YSOs is inevitably challenging, but NIRCam's resolution will greatly reduce the confusion-induced misclassification problems that plagued Spitzer observations \citep[e.g.,][]{Koepferl:2015aq,Jang:2022jw}.
    }
    \label{fig:YSOCMD}
\end{figure}

To improve the realism of the YSO modeling, we apply a set of selections to the \citet{Robitaille:2017lf} model grid that excludes unrealistic sources and classifies YSO models by their theoretical stage, allowing some distinction between their ages.
Figure \ref{fig:YSOCMD} shows a simulated population of YSOs in the selected F212N and F480M filters.
The model population is generated assuming a constant star formation rate of 0.1 $\mathrm{M}_\odot$ yr$^{-1}$, timescales of the Stage 0, I, and II evolutionary stages of 0.1, 0.5, and 5.0 Myr, respectively, and an assumed $L/\mathrm{L}_\odot=(M/\mathrm{M}_\odot)^3$ luminosity scaling.
They are sampled from a Salpeter IMF.
Only stars with $M>4$ M$_\odot$ ($L>64 \mathrm{L}_\odot$) are included, as these are roughly the YSOs expected to be dominated by stellar luminosity rather than accretion luminosity.
For lower-mass sources, the luminosity scatters between extremely faint ($<0.1 L_\odot$) and quite bright ($>10 L_\odot$) depending on their accretion rate \citep{Dunham:2014yu}.
This is a very coarse population synthesis, but it suggests that there are $\sim1500$ detectable $M>4$ M$_\odot$ Stage 0/I YSOs across the CMZ, most of which are Stage I (i.e., they have envelopes).
There should be $\sim3000$ Stage 0 and $\sim18000$ Stage I sources from 0.5-4 M$_\odot$, this time adopting a Kroupa mass function.
An unknown fraction of these sources may be detectable along with the above $>4$ M$_\odot$ YSOs if they are accreting at $\dot{M}\gtrsim10^{-5}$ M$_\odot$ yr$^{-1}$.
There may be about $10\times$ as many Stage II (disk, but no envelope) objects that may have substantial infrared excess.

With multiple epochs, we will observe variability in YSOs.
Even with multiple different bands at different epochs, large variability events like FuOr bursts will be clear.

\subsection{Motivation for MIRI channel}
\begin{figure}
    \centering
    \includegraphics[width=\textwidth]{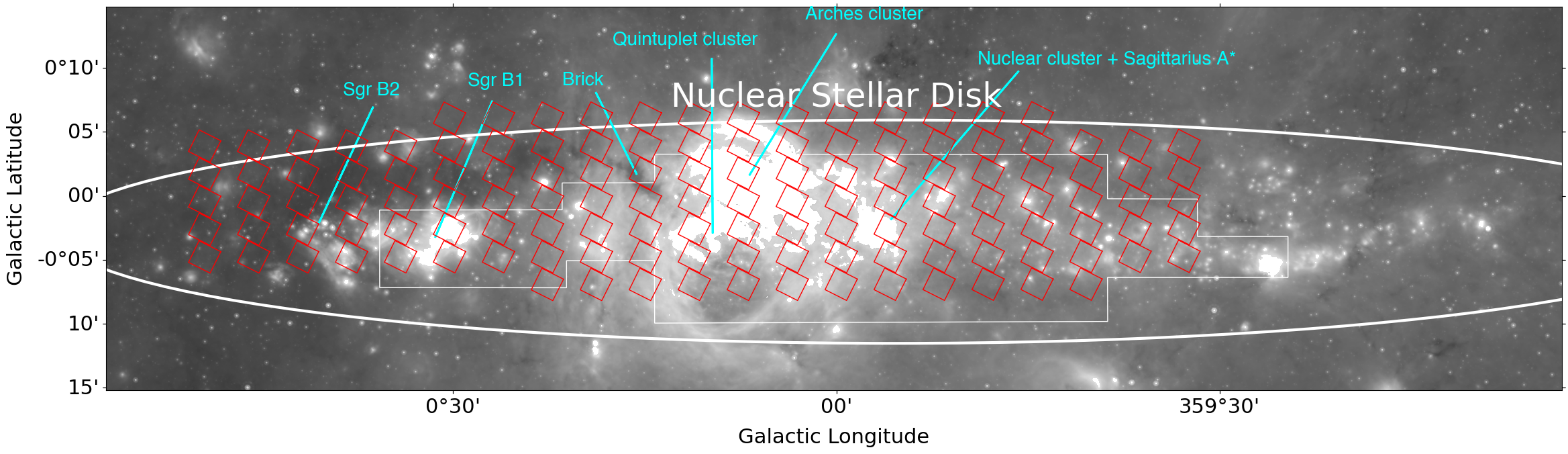}
    \caption{Overview plot like Figure \ref{fig:obsoverview} but with MIRI pointings included.
    The background image is MIPS 24 $\mu$m.
    }
    \label{fig:obswithmiri}
\end{figure}

With any NIRCam mosaic, as long as a mode other than BRIGHT1 is adopted as the readout mode, parallel observations are possible.
Any mosaic of the Galactic Center can therefore include a sparsely sampled MIRI mosaic ``for free''.
Figure \ref{fig:obswithmiri} shows where those MIRI pointings would land.
We investigate here which channel is optimal, assuming we have only one pass and incomplete coverage.

For YSO identification, the longer wavelengths are always better in the sense that they are more reliable: an excess at F2100W or F2550W would certainly indicate circumstellar dust, which is strong evidence for a YSO (or a rare class of evolved star such as a post-AGB star).
Figure \ref{fig:YSOCMD} shows the clear separation from stars, even given substantial extinction, in the F2100W band.
The longer wavelengths can also pierce through greater extinction, which is useful for identifying Class 0/I YSOs that are generally deeply embedded.
However, the longer MIRI wavelengths have relatively poor resolution that may be a significant problem for the Galactic center, since there is extended bright emission that limits point source sensitivity.
Figure \ref{fig:obswithmiri} shows that, while some pointings cross very bright 24$\mu$m emission and will likely have poor point source recovery (in the range $-10' < \ell < 20'$), the majority of the anticipated MIRI pointings cover relatively dark regions in which the benefits of the long wavelength outweigh the challenges.
While multiple epochs may allow multiple filters, we recommend prioritizing F2100W, F1280W, and F770W, in that order.

\section{Appendix: Observational Strategy Tradeoffs considered}

We summarize here discussions about alternative strategies that have been considered in preparation for this large program.

\subsection{Shorter wavelengths}
Shorter wavelengths, such as the near-infrared J and H bands, are highly desirable for stellar classification.
They are more useful than the selected K-band (F212N) for determining stellar properties.
Additionally, the higher angular resolution results in higher accuracy astrometry.

However, several considerations work against these shorter wavelengths.
First is extinction: $A_V\sim30$ translates to roughly $A_K\sim3$, but $A_J\sim6$.
As seen in Figure \ref{fig:obs}, the K-band-equivalent filter will be sensitive enough to detect the main sequence turnoff (MSTO), while additional 3 magnitudes of extinction would push it below our detection limit.
Additionally, the sensitivity of the shorter-wavelength bands to stellar properties means they are less directly useful for dereddening.
Obtaining the K-band-equivalent data will be necessary to create usable color-magnitude diagrams.
Finally, there is a purely observational concern: wavelengths short of 2 microns require sub-pixel dithers with NIRCam, which significantly adds to the total observing time and makes the overall program more difficult to execute.

\subsection{Narrow, Medium, or Broad filters}
Broadband filters are intrinsically more sensitive to starlight and less affected by narrow-band emission and absorption features.
However, broadband filters have ambiguity in their effective central wavelength that depends on spectral type \citep[e.g., Fig. 9 of ][]{Clark:2018yp}.
Furthermore, broadband filters saturate on fainter stars than medium- or narrow-band filters.

We considered both medium- and narrow-band filters.  Medium-band again saturate on fainter stars, but they have the advantage over narrowband of being less sensitive to widths of line profiles in stellar atmospheres, and they are more sensitive.

\subsection{Mosaicing and integration strategies \label{sec:mosaic}}
\begin{figure}
    \centering
    \includegraphics[width=\textwidth]{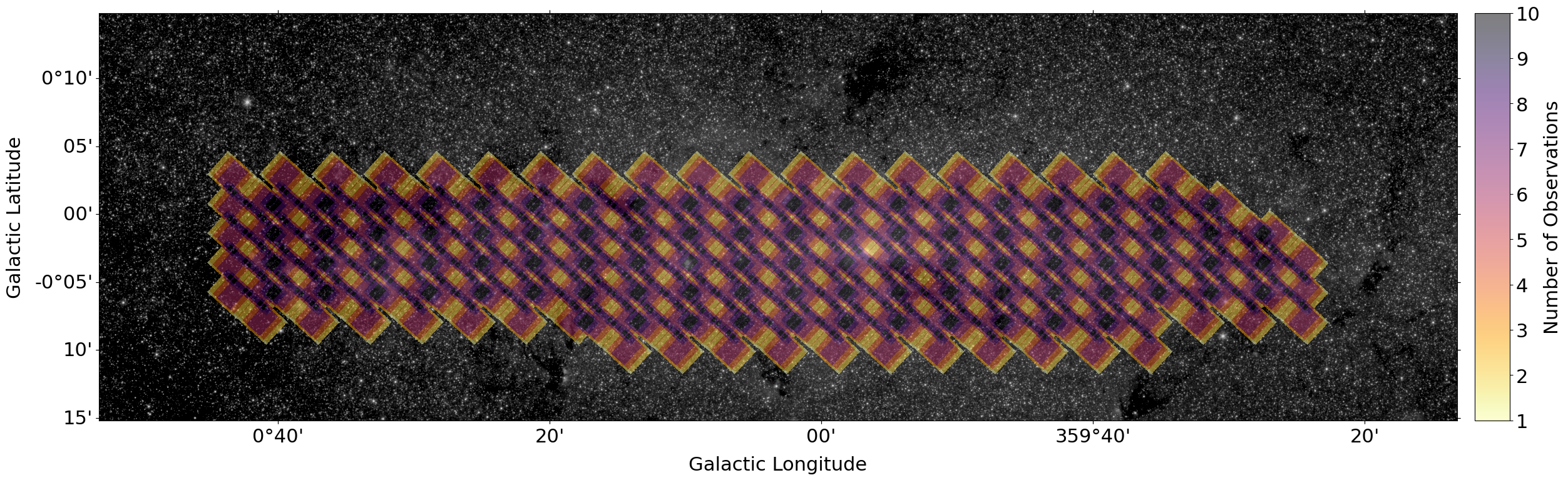}
    \includegraphics[width=\textwidth]{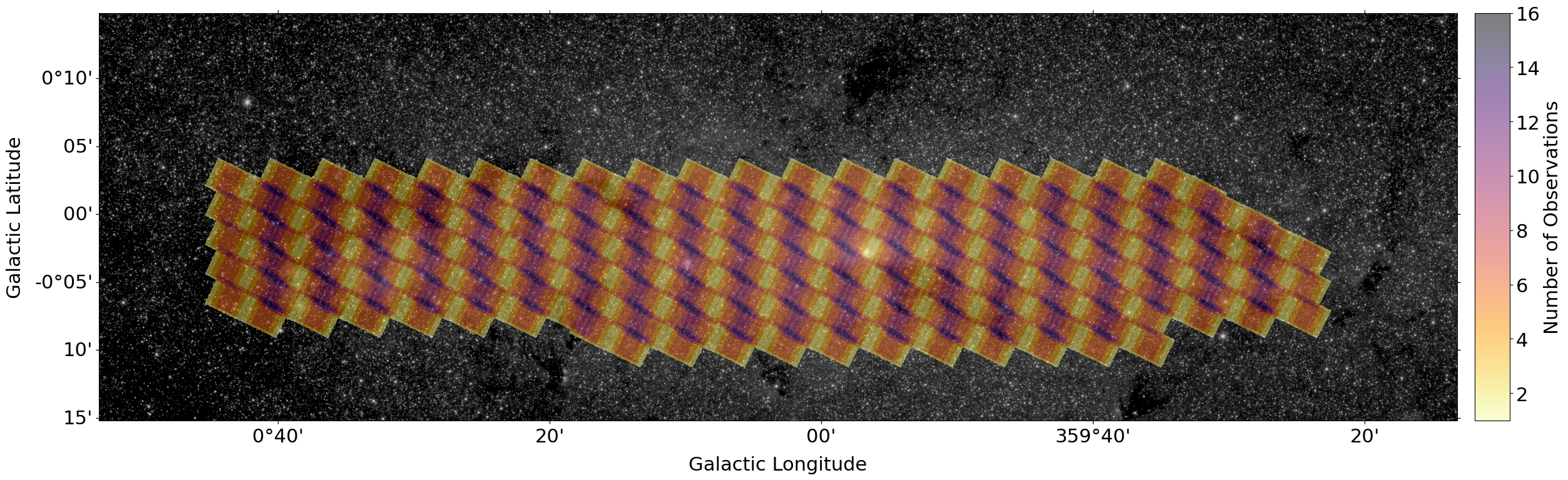}
    \caption{The NIRCam long-wavelength coverage map overlaid in color on the Spitzer 4.5$\mu$m image.
    The top and bottom panels show PA=79 and 95 degrees, respectively.
    The lower PAs have some low-coverage positions and gaps that appear toward the edge of the mosaic.
    }
    \label{fig:coveragemap}
\end{figure}

NIRCam is a very flexible instrument that can be used with a wide variety of readout and dithering strategies.
The targeted area, described in Figure \ref{fig:obsoverview}, requires 118 NIRCam pointings to fully cover.
We select a grid pattern that is guaranteed to image every pixel in the mosaic with at least 6 independent exposures; because of the range of allowed PAs (approximately 79 to 95 degrees), we require some overlap between pointings.
Figure \ref{fig:coveragemap} shows the coverage map at each possible angle; the real mosaic will end up being a mix of position angles.
With such a large number of pointings, it is important to minimize per-pointing time costs.

We investigated adopting sub-pixel dithering to improve image quality and possibly enable shorter wavelengths.
Even for the smallest sub-pixel dither strategy (2 subpixel dither positions), the overheads are substantial: for the same integration time, the observing cost is about 30\% higher.

Additionally, we explored using small-offset mosaics, which do not require new guidestar acquisition and can therefore be done in the same Visit.
Using such a strategy allows us to reduce the number of pointings to 66 and results in approximately the same total charged time (4\% greater), but with about 28\% more on-source integration time.

For readout strategies, we considered BRIGHT1, BRIGHT2, and SHALLOW2.
BRIGHT1 is optimal for preserving the flux measurements of medium-bright stars, but offers only a modest improvement over the FRAME0 read for such stars.
Since BRIGHT1 precludes parallel-mode observations, we avoid it: the value of simultaneous parallel-mode MIRI observations is too high to exclude.
SHALLOW2 observations increase the observing time substantially (35\%) without significantly improving the detection limits of the survey (since we generally hit the confusion limit rather than the sensitivity limit for source detection).
We therefore recommend BRIGHT2 + MIRI parallel observations.

\begin{figure}
    \centering
    \includegraphics[width=\textwidth]{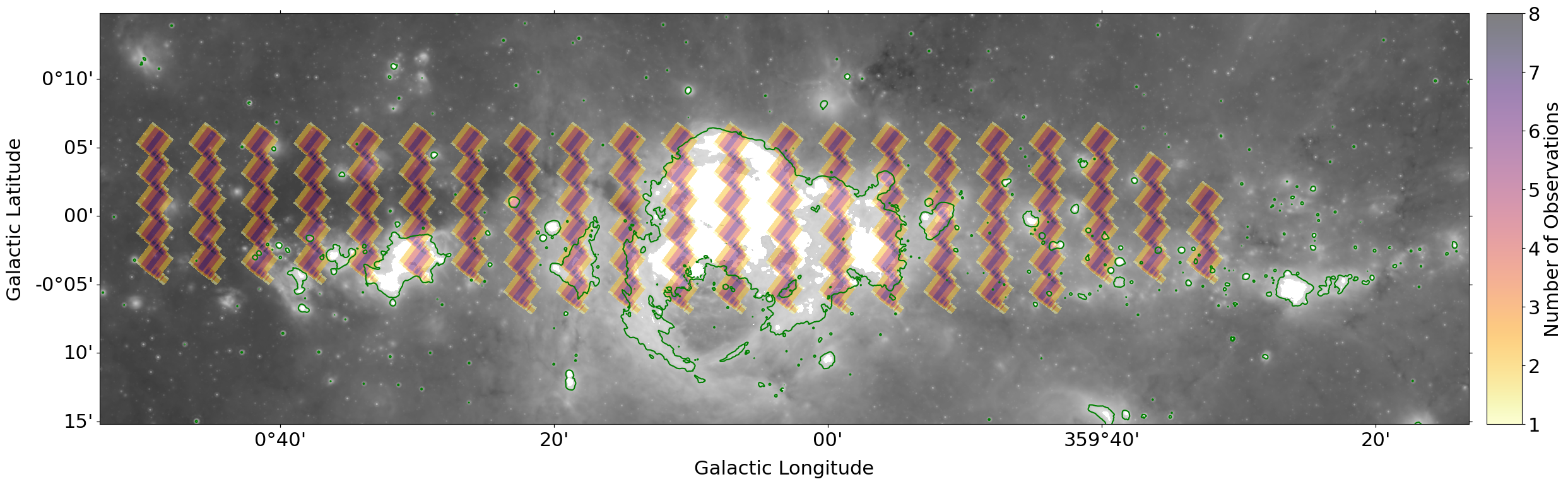}
    \includegraphics[width=\textwidth]{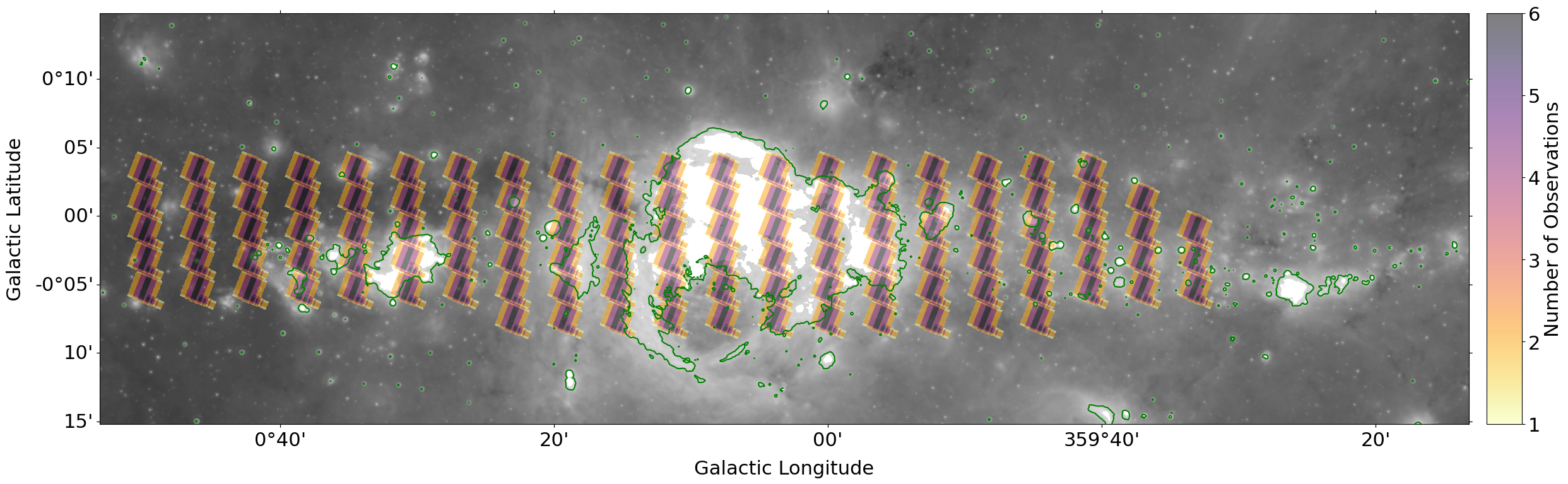}
    \caption{The MIRI coverage map overlaid in color on the Spitzer 24$\mu$m image.
    The top and bottom panels show PA=79 and 95 degrees, respectively.
    The MIRI coverage will be sparse, but will provide a fair sampling of the Galactic Center.
    The green contour shows the 1000 MJy sr$^{-1}$ threshold, which is roughly where the MIRI F2100W filter will saturate in 6-group exposures.
    Clearly, some fields will be entirely saturated, but the majority of the observed area will not.
    }
    \label{fig:miricoveragemap}
\end{figure}

\end{document}